\documentclass[]{interact}
\usepackage[numbers,sort&compress]{natbib}
\usepackage[table]{xcolor}
\usepackage{epstopdf}
\usepackage[caption=false]{subfig}

\usepackage[numbers,sort&compress]{natbib}
\bibpunct[, ]{[}{]}{,}{n}{,}{,}

\makeatletter
\def\NAT@def@citea{\def\@citea{\NAT@separator}}
\makeatother

\theoremstyle{plain}

\theoremstyle{definition}

\theoremstyle{remark}

\def\ve{{\varepsilon}}
\def\w{\omega}
\def\bk{{\bf k}}
\def\bkq{{{\bf k}+ {\bf q}}}

\def\bq{{\bf q}}

\newcommand{\ket}[1]{ | #1 \rangle }
\newcommand{\bra}[1]{ \langle #1 | }
\def\d{\delta}
\def\>{\rangle}
\def\<{\langle}
\def\D{\partial}
\def\aq{\hat a_{\bq \nu} }
\def\adq{\hat a_{\bq \nu}^{\dagger} }

\begin{document}

\title{Ultrafast dynamics of electrons and phonons: 
from the two-temperature model to the time-dependent Boltzmann equation} 

\author{
\name{Fabio Caruso\textsuperscript{a}\thanks{CONTACT F. Caruso. Email: caruso@physik.uni-kiel.de} and Dino Novko\textsuperscript{b}}
\affil{\textsuperscript{a}Institut f\"ur Theoretische Physik und Astrophysik, Christian-Albrechts-Universit\"at zu Kiel, Kiel, Germany; 
\textsuperscript{b}Institute of Physics, Zagreb, Croatia}
}

\maketitle

\begin{abstract}
The advent of pump-probe spectroscopy techniques paved the way to the exploration of the ultrafast dynamics of electrons and phonons in crystalline solids. Following photo-absorption of a pump pulse and the initial electronic thermalization, the dynamics of electronic and vibrational degrees of freedom is dominated by electron-phonon and phonon-phonon scattering processes. 
The two-temperature model (TTM)  and its generalizations -- as, e.g., the non-thermal lattice model (NLM) -- provide valuable tools to describe these phenomena and the ensuing coupled electron-phonon dynamics over timescales ranging between few hundreds of femtoseconds and tens of picoseconds. While more sophisticated theoretical approaches are nowadays available, the conceptual and computational simplicity of the TTM makes it the method of choice to model thermalization processes in pump-probe spectroscopy, 
and it keeps being widely applied in both experimental and theoretical studies. 
In the domain of ab-initio methods, the time-dependent Boltzmann equation (TDBE) ameliorates many of the shortcomings of the TTM and it enables a realistic and parameter-free description of ultrafast phenomena with full momentum resolution. 
After a pedagogical introduction to the TTM and TDBE,  in this manuscript we review their application to the description of ultrafast process in solid-state physics and materials science  as well as their theoretical foundation. 
\end{abstract}

\section{Introduction}

Electron-phonon coupling is one of the fundamental interaction mechanism in
solids \cite{GiustinoRMP}. It determines the temperature dependence of the
electron band structure \cite{Allen_1976,PhysRevB.23.1495} and of the
fundamental gap \cite{PhysRevLett.105.265501}, it influences the band effective
masses \cite{PhysRevB.97.121201} and charge transport \cite{allen71,park14,Ponce2020}, it
underpins the absorption of light in indirect band-gap materials
\cite{PhysRevB.81.241201,PhysRevLett.108.167402}, plasmon damping
\cite{brown16,novko17,PhysRevB.97.205118}, Kohn anomalies
\cite{PhysRevLett.93.185503,Lazzeri2006,calandra10,Caruso/PRL/2017,novko18,eiguren20,novko20}, the formation of
quasiparticles, as e.g. Fr\"ohlich \cite{Verdi2017,EuO,Caruso/PRB/2018} and
Holstein \cite{Kang2018,Garcia-Goiricelaya} polarons, and it underlies several exotic states of matter, such as superconducting \cite{carbotte03,luders05,marques05,margine13} and charge-density-wave \cite{calandra09,rossnagel11,zhu15} phases.
Phonon-assisted scattering processes further govern the ultrafast dynamics of
electrons and phonons and determine the time scale for the thermalization of
electronic and vibrational degrees of freedom \cite{PhysRevLett.112.257402}.
For example, a photo-excited electron distribution loses its energy by
undergoing phonon-assisted scattering processes which entails the emission and
absorption of phonons. 

Pump-probe experiments provide a versatile tool to directly probe these
phenomena \cite{lisowski_ultra-fast_2004,torre21}.  Time- and angle-resolved
photoemission spectroscopy, for example, enables to directly probe the dynamics
of photoexcited electrons and holes with energy and momentum resolution
\cite{bovensiepen12,rohde_ultrafast_2018}. Additionally, pump-probe scattering techniques as,
e.g., ultrafast electron diffuse scattering complement optical and
photoemission measurements by providing direct insight into the dynamics of the
crystalline lattice and electron-phonon scattering processes with time and
momentum resolution, and they are suitable to attain a detailed understanding
of the energy flow between hot electrons and the crystalline lattice
\cite{waldecker_eph_2016,Stern2018,durr21}.

The two-temperature model (TTM) \cite{kaganov57,lifshits60,anisimov73,Allen1987}
describes the dynamics of electron and phonons as the thermalization involving
two coupled thermal reservoirs.  This idea provides perhaps the simplest and
most intuitive description of the thermalization of electronic and vibrational
degrees of freedom in systems out of equilibrium and it has seen wide
application to the description of ultrafast processes in solids
\cite{corkum_thermal_1988,brorson_femtosecond_1990,rethfeld_ultrafast_2002,jiang_improved_2005,carpene_ultrafast_2006,lin_electron-phonon_2008,conforti_derivation_2012,mueller_relaxation_2013,wilson_two-channel_2013,shin_extended_2015,waldecker_eph_2016,brown_ab_2016,loncaric16,xian_analytic_2017,an_generalized_2017,Darancet,maldonado_theory_2017,maldonado_tracking_2020,novko_ultrafast_2019,smirnov_copper_2020,naldo_understanding_2020,pietro20,ritzmann_theory_2020,miao_nonequilibrium_2021,novko_first-principles_2021,sidiropoulos21}.
The TTM has been employed to examine the fingerprints of ultrafast dynamics in
several experimental techniques including photoemission spectroscopy
\cite{fujimoto_femtosecond_1984,fann_direct_1992,fann_electron_1992,hertel_ultrafast_1996,perfetti07,bonn_ultrafast_2000,johannsen_direct_2013,gierz_snapshots_2013,yang17,caruso_photoemission_2020},
Raman scattering \cite{novko_ultrafast_2020,pellatz21}, ultrafast electron diffuse
scattering
\cite{chase_ultrafast_2016,waldecker_eph_2016,rene_de_cotret_time-_2019,maldonado_tracking_2020},
and pump-probe optical techniques
\cite{elsayed-ali_time-resolved_1987,schoenlein_femtosecond_1987,brorson_femtosecond_1990,groeneveld_effect_1992,juhasz_direct_1993,sun_femtosecond-tunable_1994,hohlfeld_time-resolved_1997,wellershoff_role_1999,hohlfeld_electron_2000,bonn_ultrafast_2000,hase_ultrafast_2005,della_valle_real-time_2012,ortolani_pump-probe_2019,novko2019,bresson_improved_2020,obergfell20,chan21,chang_electron_2021}.
It has further been extended to account for the emergence of coherent phonons
\cite{tang_coherent_2008,giret_entropy_2011}, ultrafast dynamics in warm-dense
matter
\cite{an_generalized_2017,bresson_improved_2020,holst_ab_2014,leguay_ultrafast_2013,petrov_modeling_2021,simoni_first-principles_2019}, thermalization via electron-plasmon channel \cite{benhayoun21},
spin dynamics \cite{beaurepaire_ultrafast_1996}, magnon scattering with
lattice degrees of freedom
\cite{sanders_effect_1977,schreier_magnon_2013,agrawal_direct_2013,montagnese_phonon-magnon_2013,hohensee_magnon-phonon_2014,liao_generalized_2014}, and thermalization of magnetic-nematic order \cite{patz14}.
An effort has been also made in upgrading the TTM to account for the nascent non-equilibrium electron distribution, while retaining the numerical simplicity of the model \cite{sun_femtosecond-tunable_1994,carpene_ultrafast_2006,tsibidis18,uehlein2021}.
{While the application of the TTM to ultrafast electron dynamics in metals and laser ablation has been reviewed in Refs.~\cite{singh10} and \cite{rethfeld17}, respectively, we here focus on its application to coupled electron-phonon dynamics in the context of first-principles simulations.}
Furthermore, the non-thermal lattice model (NLM), also referred to as $N$-temperature or multi-temperature model, extends the TTM
to account for anisotropies in electron-phonon interaction \cite{waldecker_eph_2016,waldecker17,maldonado_theory_2017,lu18,ono18}. In this framework,
the lattice is treated as a set of $N$ distinct thermal reservoirs coupled to the
electrons.  At variance with the TTM, this generalization enables to account
for the establishment of  {\it hot phonons}  or, more generally, a non-thermal
state of the lattice, characterized by the enhanced population of the phonon
modes which actively take part in the hot-carrier relaxation
\cite{kampfrath_strongly_2005,perfetti07,ishida11,dalconte12,johannsen_direct_2013,mansart13,novko_ultrafast_2020}.

A rigorous description of electron-phonon coupling and its influence on the
dynamics inevitably requires to account for phase-space constraints in the
phonon-assisted scattering processes as well as for the anisotropies of the
electron-phonon coupling. A suitable approach to attain this requirement is the
time-dependent Boltzmann equation (TDBE). 
Ultrafast dynamics simulations based on the TDBE formalism have been widely
employed to described non-equilibrium processes involving electrons and phonons
in condensed matter.  The TDBE has been applied to investigate the
thermalization of electrons and phonons following photoirradiation in metals
\cite{pines62,Grinberg1990,Gusev1998,Rethfeld2002,kim11,Mueller_2013b,baranov14,Marini2019,Ono2020},
semiconductors \cite{COLLET1986153,Binder1992,Bernardi2017,Perturbo}, and
two-dimensional
\cite{Knorr2009,Tani2012,brida_ultrafast_2013,kratzer_relaxation_2019,tong_toward_2021,caruso/2021}
and layered  \cite{seiler_NL_2021} materials, and it  has further been extended
to investigate the ultrafast magnetization dynamics in photo-excited
ferromagnets \cite{Mueller_2011,Mueller_2013}.  In comparison to simplified
models -- as, e.g., the TTM and NLM -- a major advantage of the TDBE approach
consists in enabling the investigation of the coupled non-equilibrium dynamics
of electrons and phonon populations with a full momentum resolution. This aspect
is particularly important to capture the anisotropic population of the
electronic and vibrational states in reciprocal space which may be established
out of equilibrium.  For example, the absorption of circularly polarized light
in transition metal dichalcogenide monolayers is governed by valley-selective
circular dichroism
\cite{cao_valley-selective_2012,mak_control_2012,zeng_valley_2012}: carriers
are photoexcited in either the K or $\overline{\rm K}$ valleys in the Brillouin
zone depending on the light helicity, leading to an anisotropic electronic
distribution which cannot be captured via a density-of-state approximations
\cite{Caruso2022}.  Similarly, highly anisotropic phonon populations in
reciprocal space can be established upon the preferential emission of
strongly-coupled phonons \cite{caruso/2021,tong_toward_2021}, leading to the
characteristic spectral signatures in ultrafast diffuse scattering techniques
\cite{rene_de_cotret_time-_2019,seiler_NL_2021,otto_mechanisms_2021,PhysRevB.104.205109,PhysRevLett.127.207401}.

This manuscript aims at introducing the fundamentals of model approaches and of
the time-dependent Boltzmann equation in the context of the non-equilibrium
dynamics of coupled electrons and phonons.  In particular, in
Sec.~\ref{sec:model} we introduce the TTM and in Sec.~\ref{sec:NLM} its
generalization to several temperature reservoirs, i.e., to the NLM.  In
Sec.~\ref{sec:bte}, we introduce the TDBE and discuss its theoretical
foundation (Sec.~\ref{sec:bte_deriv}) as well as its relation to the TTM.  In
Sec.~\ref{sec:graphene}, after a concise overview of the experimental
signatures of  non-equilibrium processes in graphene (Sec.~\ref{sec:Ca}) we
proceed to discuss the theoretical description of the ultrafast carrier and
lattice dynamics based on the NLM (Sec.~\ref{sec:Cb}) and the TDBE
(Sec.~\ref{sec:Cc}) approaches.  Summary and outlook are discussed in
Sec.~\ref{sec:conc}. 

\section{The two-temperature model}\label{sec:model}

\begin{figure}[t]
\centering
\resizebox*{9cm}{!}{\includegraphics{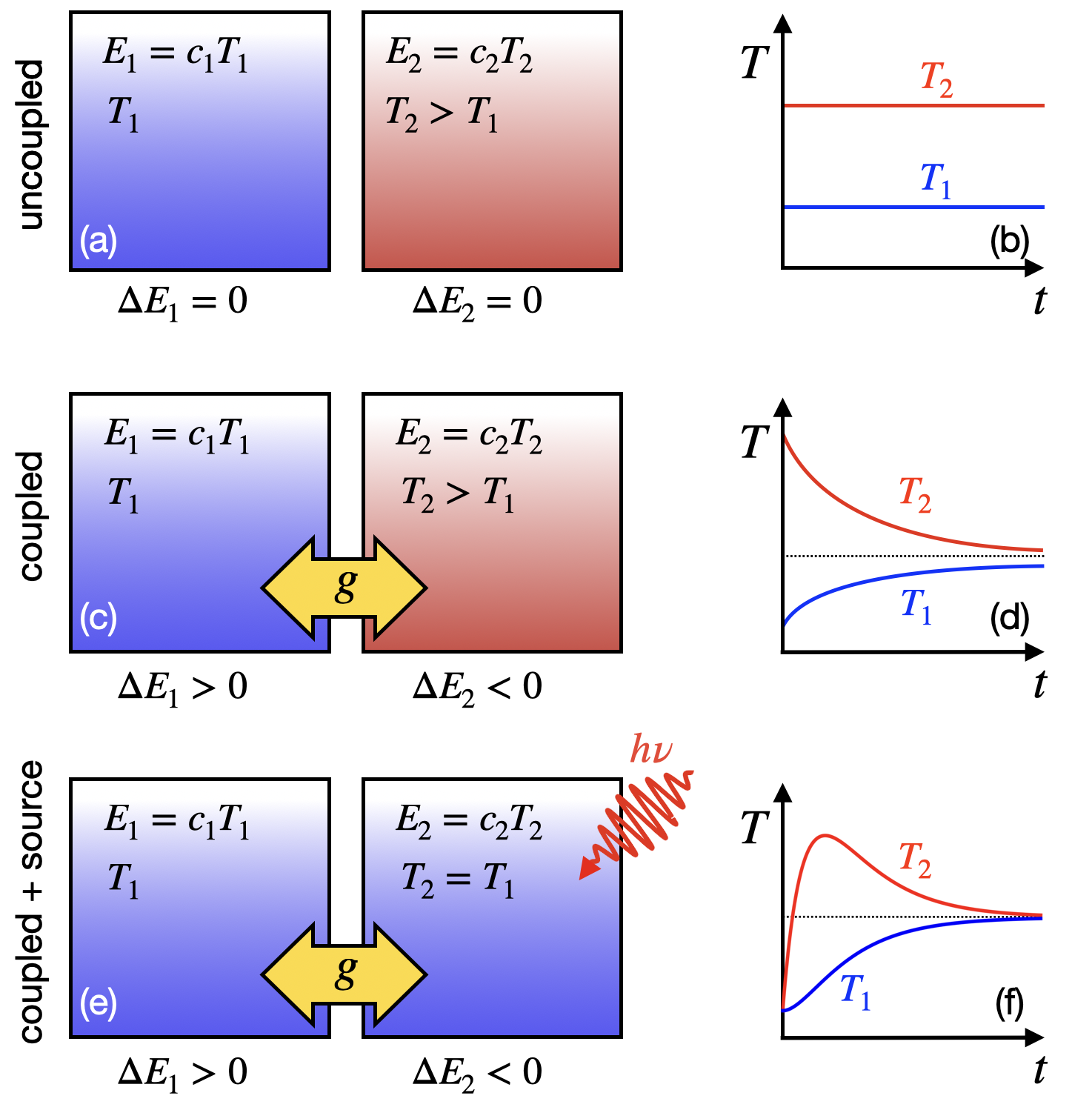}}
\caption{
Schematic representation of two thermal reservoirs at temperatures $T_1$ and
$T_2>T_1$ and energies $E_1 = C_1 T_1$ and $E_2=C_2 T_2$ --  where $C_1$ and
$C_2$ denote the heat capacities -- in absence of interactions (a), in presence
of mutual interactions characterized by a coupling constant $g$ (c), and in
presence of an external field (e).  (b), (d), and (f): Time dependence of the
temperature for the systems in (a), (c), and (e), respectively, as obtained
from the solution of the TTM.  } 
\label{fig:sketch1}
\end{figure}

The underlying ideas of the TTM \cite{kaganov57,lifshits60,anisimov73,Allen1987} are most
easily illustrated by considering the thermalization dynamics of two systems
$S_1$ and $S_2$. Quantum mechanics is not required at this point; $S_1$, $S_2$,
and their interactions can be assumed to be governed by the laws of
thermodynamics. 
If $S_1$  and $S_2$ are initially at thermal equilibrium at the temperatures
$T_1$ and $T_2$, respectively, in absence of interactions
[Fig.~\ref{fig:sketch1}~(a)] the temperature of each subsystem will remain
unaltered through time [Fig.~\ref{fig:sketch1}~(b)].  If heat can be exchanged
[Fig.~\ref{fig:sketch1}~(c)], on the other hand,  one can expect that
for a sufficiently small temperature difference $T_2-T_1$
the energy transferred from $S_1$ to $S_2$ (from $S_2$  to $S_1$) during the
interval $\Delta t$ will be linear in $T_2-T_1$ ($T_1-T_2$), leading to \cite{kaganov57}:
\begin{align}
\frac{\Delta E_1}{\Delta t} &= g_1 (T_2-T_1)\quad, \\
\frac{\Delta E_2}{\Delta t} &= g_2 (T_1-T_2)\quad .
\end{align} Here, the energy of $S_1$ is given by $E_1 = c_1 T_1$, where $c_1$
is the specific heat, and  similarly for $S_2$. $g_1$ and $g_2$ are
proportionality constants with units of $\mathrm{\left[energy\times(temperature\times time)^{-1}\right]}$.
Energy conservation throughout the thermalization dynamics requires $\Delta E_1=
-\Delta E_2$, which in turn promptly leads to condition $g_1=g_2 = g$.  In the
limit of infinitesimal time interval, the thermalization dynamics of $S_1$ and
$S_2$ can thus be formulated by the coupled first-order differential equations:
\begin{align}\label{eq:mod0a} \frac{ \D T_1}{\D t} &= \frac{g}{c_1} (T_2-T_1) \quad,\\
\frac{\D T_2}{\D t} &= \frac{g}{c_2} (T_1-T_2)\quad.\label{eq:mod0b} \end{align}
Equations \eqref{eq:mod0a} and \eqref{eq:mod0b} are the central equations of the
two-temperature model. If the temperature dependence of $c_1$ and $c_2$ are
neglected, Eqs. \eqref{eq:mod0a} and \eqref{eq:mod0b} admit analytical solutions
in the form of decaying exponentials (see Appendix). A detailed discussion of
the analytical solution of the TTM can be found in
Ref.~\cite{xian_analytic_2017}. The resulting time dependence of the
temperatures $T_1$ and $T_2$ is illustrated in Fig.~\ref{fig:sketch1}~(d). In
short, the presence of a coupling constant $g$ tends to restore a regime of
thermal equilibrium where $T_2=T_1$.  More generally, owing to the non-trivial
temperature dependence of the specific heat,  the TTM must be solved numerically
via time-stepping algorithms (e.g., the Euler or Runge-Kutta algorithms). 

The TTM is often modified to introduce a time-dependent source (driving) term
coupled to $S_1$ or $S_2$ \cite{j_k_chen_j_e_beraun_numerical_2001,lin_electron-phonon_2008}. This scenario is clearly reminiscent of pump-probe
experiments [Fig.~\ref{fig:sketch1}~(e)] whereby either electrons or lattice are
driven out of equilibrium through the coupling to ultrashort pulses. If a
coupling in the form $S(t)=\alpha e^{-\frac{|t| }{\tau}}\theta(t)$ is added to
Eq.~\eqref{eq:mod0b}, i.e., a laser pulse with the amplitude of $\alpha$ that lasts for time period of $\tau$ and starts at $t=0$, the model still admits analytic solution
(Appendix), leading to the trend illustrated in Fig.~\ref{fig:sketch1}~(f).
For a more detailed discussion on the time-dependent source term and its different functional forms we refer the reader to the Refs. \cite{rethfeld17,carpene_ultrafast_2006,maldonado_theory_2017}.
The initial increase of $T_2$ reflects the raise of temperature of $S_2$ induced by
the interaction with a source, whereas at a later stage, the thermalization
follows a similar trend to that of Fig.~\ref{fig:sketch1}~(d). 

The TTM can be straightforwardly employed to model the coupled electron-phonon
phonon dynamics in condensed matter via the following steps \cite{Allen1987,lin_electron-phonon_2008}: one of the
subsystems ($S_1$) is identified with lattice, whereas the other ($S_2$) with
the electrons. The temperatures $T_1$ and $T_2$ are identified with the effective 
temperatures of the lattice ($T_{\rm ph}$) and electrons ($T_{\rm el}$) 
and the source term $S(t)$ is employed to model the coupling to
an external light source. $c_1$ and $c_2$ are replaced by the phonon ($C_{\rm
ph}$) and electron ($C_{\rm el}$) heat capacities, which can be expressed as \cite{lin_electron-phonon_2008}: 
\begin{align}
C_{\rm el} (T) &= \int_\infty^{\infty} d\ve D_{\rm el}(\ve) \ve 
\frac{\D f (\mu,\ve,T_{\rm el})}{\D T_{\rm el}} \label{eq:dosel}\quad,\\
C_{\rm ph} (T) &= \int_0^{\infty} d\omega D_{\rm ph}(\omega) \omega
\frac{\D n (\omega,T_{\rm ph})}{\D T_{\rm ph}} \label{eq:dosph}\quad,
\end{align}
where $D_{\rm el}$ and $D_{\rm ph}$ are 
the electron and phonon density of states. 
Similarly, the coupling constant $g$ can be expressed as \cite{lin_electron-phonon_2008,GiustinoRMP}: 
\begin{align} g=\frac{\pi k_B}{\hbar D_{\rm 
el}(\varepsilon_F)}\lambda\left\langle \omega^2
\right\rangle\int_{-\infty}^{\infty} d\varepsilon D_{\rm
el}^2(\varepsilon)\left(- \frac{\partial f(\mu,\ve,T_{\rm el})}{\partial
\ve} \right), \label{eq:eq7} 
\end{align}
where the thermalization rate of electron-lattice system is governed by the electron-phonon coupling strength and second moment of the phonon spectrum, which are related to the Eliashberg function $\alpha^2F(\Omega)$ as $\lambda\left\langle \omega^2 \right\rangle = 2\int d\Omega \Omega
\alpha^2F(\Omega)$. 
The equations of the TTM, i.e., Eqs.~\eqref{eq:mod0a} and \eqref{eq:mod0b}, can thus be expressed as:
 \begin{align}\label{eq:ttm_2a}
 \frac{\D T_{\rm ph}}{\D t} &= \frac{g}{C_{\rm ph}}( { T_{\rm el}- T_{\rm ph} })\quad,  \\ 
 \frac{ \D T_{\rm el}}{\D t } 
&= \frac{g} {C_{\rm el}}  ({ T_{\rm ph}- T_{\rm el}})  
\label{eq:ttm_2b} + S(t)\quad.
\end{align}

The theoretical foundation of the TTM, and its relation to the TDBE is discussed 
in Sec.~\ref{sec:bte-ttm}.
While the electron and phonon heat capacities ${C_{\rm el}}$  and ${C_{\rm
ph}}$  can be immediately obtained from calculations based on
density-functional theory and density-functional perturbation theory \cite{baroni01},
respectively, the parameters $g$ can be estimated from first-principles
calculations of the Eliashberg function via well-established simulation
packages \cite{Ponce2016}. This procedure enables the solution of the TTM
entirely ab-initio, without resorting to free parameters 
\cite{lin_electron-phonon_2008,brown_ab_2016,novko_ultrafast_2019,caruso_photoemission_2020,novko_first-principles_2021}.
Alternatively, the $g$ can be deduced by experimental data, e.g., by fitting Eq.~\eqref{eq:ttm_2b} to pump-probe photoemission measurements (see, e.g., Sec.~\ref{sec:graphene}) \cite{johannsen_direct_2013}. 

As discussed in Sec.~\ref{sec:bte-ttm}, 
the application of the TTM to the non-equilibrium dynamics of
electrons and phonons in solids can be justified through its formal
derivation from the time-dependent Boltzmann equation \cite{Allen1987}. 
The description of ultrafast processes via Eqs.~\eqref{eq:mod0a} and 
\eqref{eq:mod0b}, however, entails two main approximations: 
(i) at each time steps throughout the dynamics, electrons are assumed 
to populate electronic bands according to a Fermi-Dirac function at the
effective temperature $T_{\rm el}$; (ii) the lattice is assumed to be at
thermal equilibrium throughout the dynamics, i.e., all bosonic occupations are
described by the Bose-Einstein statistics at the effective temperature $T_{\rm ph}$. 
These approximations limit the domain of applicability of the TTM. 
Because of the approximation (i), the TTM is unsuitable to describe the 
early stages of the electron dynamics ($t<100$~fs), which can be characterized 
by population inversion, the anisotropic excitation of electron-hole pairs 
in the Brillouin zone, and electron-electron scatterings. 
The domain of application of the TTM is thus restricted 
to metals, semimetals, and doped semiconductors with short electron thermalizaton times, since, on the other hand, electronic excitations in gapped systems (e.g., semiconductors) are inherently linked to 
a regime of population inversion that cannot be properly modeled via a Fermi-Dirac function. 
Additionally, the approximation (ii) makes the TTM unsuitable to 
describe the non-equilibrium dynamics of the lattice. 
The TTM is sometimes extended to model the population inversion and other forms of nascent non-equilibrium distributions by defining separately electron and hole thermal baths \cite{wang_ultrafast_2010}, i.e., electron and hole temperatures\,\cite{breusing_ultrafast_2009}, or by dividing the electronic bath into a majority of thermal and a small portion of non-thermal carriers \cite{sun_femtosecond-tunable_1994,carpene_ultrafast_2006,tsibidis18,uehlein2021}. However, in these extensions, the issue of thermalized lattice bath (ii) is still present.
In the following, we discuss how this limitation can be overcome 
by extending the TTM to account for anisotropic coupling to different phonon modes.

\subsection{The non-thermal lattice model}\label{sec:NLM}

\begin{figure}[t]
\centering
\resizebox*{12cm}{!}{\includegraphics{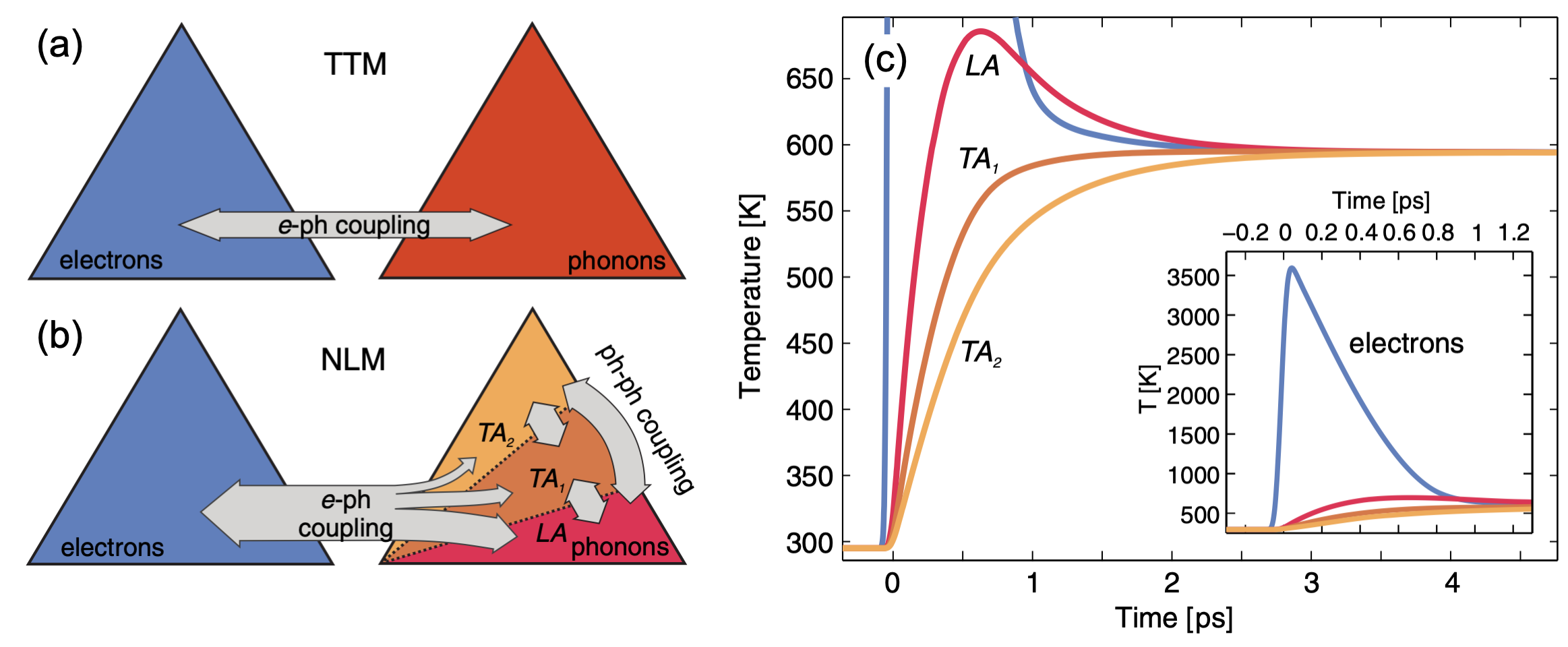}}
\caption{(a) Schematic illustration of the TTM and (b) NLM. At variance with the TTM, the NLM accounts for coupling with different subsets of phonon modes, as well as phonon-phonon interactions. 
(c) Time-dependence of the electronic (blue) and vibrational temperatures 
of the longitudinal (LA) and transverse acoustic phonons (TA$_1$ and TA$_2$)  of aluminum. 
Reproduced from Ref.~\cite{waldecker_eph_2016}. 
} 
\label{fig:sketch2}
\end{figure}

Ultrafast diffuse-scattering experiments and first-principles calculations
provide strong evidence that non-thermal regimes of the lattice -- i.e.,
vibrational states characterized by bosonic occupations which deviate
significantly from the Bose-Einstein statistics -- can be established upon
photo-excitation in both semiconducting and metallic layered compounds, such
as, black phosphorus \cite{seiler_NL_2021}, MoS$_2$ \cite{caruso/2021},
graphite \cite{rene_de_cotret_time-_2019}, graphene \cite{tong_toward_2021},
and TiS$_2$ \cite{otto_mechanisms_2021}.  Even for simple metals such as Al,
the anisotropic coupling between electrons and acoustic phonons can trigger the
emergence of a non-equilibrium vibrational states persisting for several
picoseconds \cite{waldecker_eph_2016}. 
Generally, whenever the electron-phonon interaction is dominated by one or
several strongly-coupled modes, these lattice vibrations may provide a
preferential decay channel for the relaxation of photo-excited electrons and
holes \cite{caruso_photoemission_2020}. As mentioned in the introduction, such a scenario can lead to the
formation of {\it hot phonons}, i.e., a non-thermal state of
the lattice \cite{perfetti07,dalconte12,johannsen_direct_2013,mansart13,novko_ultrafast_2020}. Because of the assumption that the lattice can be
described by a Bose-Einstein distribution at temperature $T_{\rm ph}$, the TTM
is unsuitable to describe these  phenomena
\cite{waldecker_eph_2016,rene_de_cotret_time-_2019,seiler_NL_2021}

To enable the description of hot phonons and non-thermal states of the lattice,
a generalization of the TTM to account for anisotropies in the coupling with
different subsets of lattice vibrations -- referred to as non-thermal lattice
model (NLM) \cite{waldecker_eph_2016,maldonado_theory_2017,lu18,ono18} or three-temperature model
\cite{caruso_photoemission_2020,novko_ultrafast_2020}, depending on the level of approximation -- has recently been
proposed. 
In short, while in the TTM the electrons are coupled to the 
lattice via a single coupling constant $g$ [Fig.~\ref{fig:sketch2}~(a)], 
in the NLM the lattice is partitioned into phonon groups 
characterized by distinct coupling constants $g_\mu$ [Fig.~\ref{fig:sketch2}~(b)].
Phonons characterized by higher coupling play a primary role in the 
electronic relaxation, and therefore exhibit a larger effective temperature 
on short timescales, whereas on longer timescales thermal equilibrium 
is restored by phonon-phonon coupling.  
This behaviour is schematically illustrated in
Fig.~\ref{fig:sketch2}~(c) for aluminum, where only three acoustic phonons are
present.

In general, by partitioning the $N_{\rm ph}$ normal modes of the lattice into
$N_{g}$ groups, where $1 < N_g \leq N_{\rm ph}$ (for $N_g=1$, the TTM is
recovered), the total energy can be expressed as 
$E_{\rm ph} =
\sum_{\mu=1}^{N_{\rm g}}  E_{\rm ph}^{\mu}$. 
The rate of change of the energy of the
$\mu$-th phonon group can be expressed as $\D_t E^\mu_{\rm ph} = C^\mu_{\rm ph}
\D_t T^\mu_{\rm ph}$, where ${C^\mu_{\rm ph}}$ is the specific heat of the
$\mu$-th phonon group, which is obtained by replacing the group density of
states in Eq.~\eqref{eq:dosph}. 
The NLM can thus be formulated as a set of coupled first-order differential equations which relates the
electronic temperature $T_{\rm el}$ to the temperatures $T_{\rm ph}^\mu$ of the
$N_g$ phonon groups: 
\begin{align}\label{eq:NLMa} \frac{\D T^\mu_{\rm ph}}{\D
t} &= \frac{g_\mu}{C^\mu_{\rm ph}}( { T_{\rm el}- T^\mu_{\rm ph} }) +
\sum_{\mu'} \frac{T^{\mu'}_{\rm ph}- T^{\mu}_{\rm ph}}{\tau_{\mu\mu'}} \quad,
\\
\frac{ \D T_{\rm el}}{\D t } &=  \sum_\mu^{N_{g}} \frac{g_\mu}{C_{\rm el}}  ({
T^\mu_{\rm ph}- T_{\rm el}}) + S(t) \label{eq:NLMb}\quad.
\end{align} 
Here, the coupling constant  $g_\mu$
is defined as in Eq.~\eqref{eq:eq7} by restricting the sum to all phonons in the
$\mu$-th group,  
 ${\tau_{\mu\mu'}}$ denotes the time constant for the decay of
the $\mu$-th subgroup due to phonon-phonon interaction with phonons in the
$\mu'$-th group, and it can be obtained from first-principles via calculations
of the phonon-phonon scattering matrix elements \cite{maldonado_theory_2017}.

As illustrated in Fig.~\ref{fig:sketch2}~(c) for Al
\cite{waldecker_eph_2016}, at variance with TTM, the NLM enables to
account for the establishment of a non-thermal regime in the lattice. In
particular, following photo-excitation of the electrons, energy is transferred
to each phonon group $\mu$ proportionally to the coupling constant $g_\mu$,
thus, leading to a discrepancy among the vibrational temperatures $T^\mu_{\rm
ph}$. On longer timescales, the onset of phonon-phonon scattering (via the
second term in Eq.~\eqref{eq:NLMa}), drives the lattice back to a thermalized
regime, where all vibrational temperatures coincides. 

With the exception of elemental metals,  where the 
identification of different phonon groups is straightforward
due to the reduced vibrational degrees of freedoms 
(see, e.g., Fig.~\ref{fig:sketch2}), in compounds with several 
atoms in the unit cell the definition of phonon groups is 
to some extent arbitrary, and it represents a shortcoming of the NLM. 
In some limiting cases (e.g., cuprates, graphene, and MgB$_2$ \cite{perfetti07,caruso_photoemission_2020,novko_ultrafast_2020}), 
it is possible to distinguish strongly and weakly coupled modes and, 
correspondingly, phonons can be grouped according to their coupling strength. 
The arbitrariness in the definition of phonon groups can be 
lifted by resorting to a first-principles description of the 
ultrafast dynamics of electrons and phonons based on the time-dependent Boltzmann equation.

\section{The time-dependent Boltzmann equation} \label{sec:bte}
\begin{figure}[t]
\begin{center}
   \includegraphics[width=0.98\textwidth]{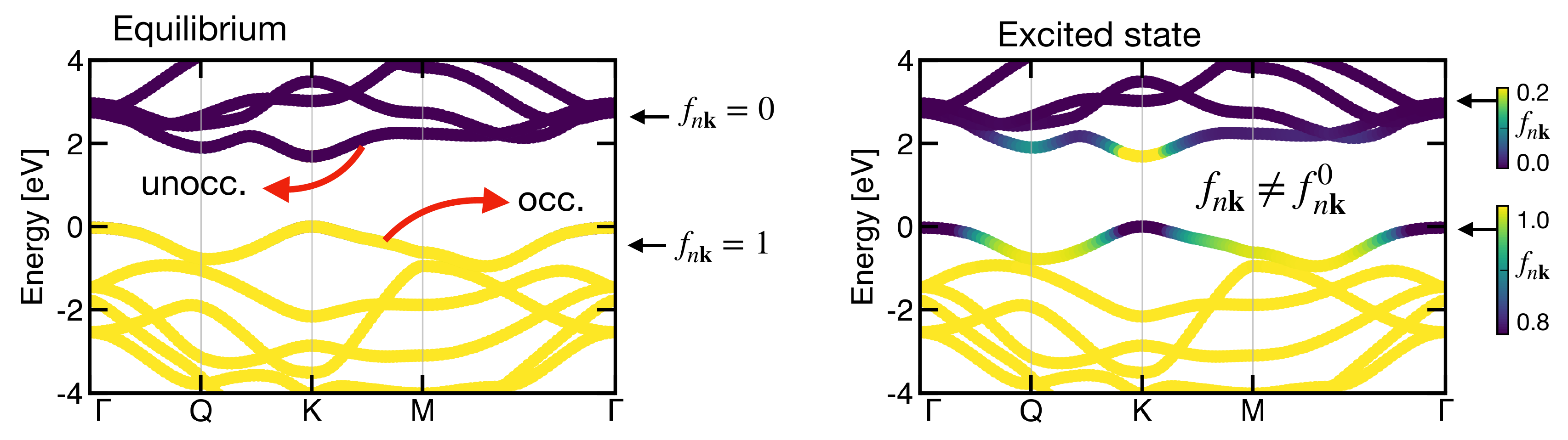}
\caption{\label{fig:car21}
Electron distribution function $f_{n\bk}$ superimposed to the band structure of monolayer MoS$_2$. Energies are relative to the Fermi level. At equilibrium (left), bands are occupied according to the Fermi-Dirac statistics (Eq.~\eqref{eq:fd}). 
Adapted from Ref.~\cite{caruso/2021}.
}
\end{center}
\end{figure}
The time-dependent Boltzmann equation (TDBE) constitutes an optimal compromise between
accuracy and efficiency to investigate the ultrafast dynamics of coupled
electron-phonon systems. In the TDBE,  
the dynamics of electronic and vibrational excitations are described 
by changes of the electron and phonon distribution functions 
$f_{n\bk}(t)$ and $n_{\bq\nu}(t)$, respectively, whereas  
electron and phonon energies are left unchanged throughout the dynamics.
At thermal equilibrium, $f_{n\bk}$ and $n_{\bq\nu}$
are time independent and they coincide with the Fermi-Dirac and the
Bose-Einstein occupations $f^{\rm 0}_{n\bk}$ and $n^{\rm 0}_{\bq\nu}$: 
\begin{align} \label{eq:fd}
f^{\rm 0}_{n\bk}(T) &= \left[ e^{ {(\ve_{n\bk}-\ve_{\rm F})}/{k_{\rm B}T}} + 1 \right]^{-1}\quad, \\
n^{\rm 0}_{\bq\nu}(T) &=\left[ e^{ \hbar\omega_{\bq\nu}/{k_{\rm B}T}} - 1 \right]^{-1}.
\end{align}
Here, $\ve_{\rm F}$  is the Fermi energy, 
$\ve_{n\bk}$ is the single-particle energy of a Bloch electron, and $\hbar\omega_{\bq\nu}$ the phonon energy. 
This case is exemplified by the left panel of Fig.~\ref{fig:car21}, where 
the Fermi-Dirac occupations are superimposed to the band structure of monolayer MoS$_2$, with yellow (blue) denoting fully 
occupied (empty) states with $f_{n\bk}=1$ ($f_{n\bk}=0$).
In this framework, a regime of non-equilibrium requires either $f_{n\bk}$ or $n_{\bq\nu}$ (or both) 
to differ from the equilibrium Fermi-Dirac and the Bose-Einstein occupations, as illustrated 
in the right panel of Fig.~\ref{fig:car21}. 
The non-equilibrium distributions change over time, and their dynamics is determined by the TDBE: 
\begin{align}
\label{eq:bte_f0}  {\D_t f_{n\bk}(t)}     &= \Gamma_{n{\bk}}(t) 
\\
\label{eq:bte_n0}  {\D_t n_{\bq \nu}(t)}  &= \Gamma_{{\bq\nu}}(t)\quad,
\end{align}
where $\D_t = \D / \D t$ and $\Gamma_{{n\bk}}$ and 
$\Gamma_{{{\bq}\nu}}$ denote the collision integrals for electrons and phonons. 
The numerical solution of Eqs.~\eqref{eq:bte_f0} and \eqref{eq:bte_n0} 
 requires the development of  suitable approximations for the evaluation of the collision integrals.  
In short,  $\Gamma_{{n\bk}}$ and $\Gamma_{{{\bq}\nu}}$ account for the 
several scattering mechanisms which may lead to changes of the 
distributions functions as, e.g.,  electron-electron, electron-phonon, 
phonon-phonon, and impurity scattering as well as the coupling to 
external fields. 
The recent development of electronic structure codes for the study of
electron-phonon and phonon-phonon coupling has enabled to estimate
the contribution of these scattering processes to
collision integrals, enabling the investigation of the coupled
electron-phonon dynamics entirely from first principles \cite{PhysRevLett.112.257402,Darancet,Ono2020,caruso/2021,tong_toward_2021}.

\subsection{First-principles expressions for the collision integrals}\label{sec:bte_deriv}
 
In the following, we outline the derivation of the 
collision integral due to the electron-phonon and phonon-phonon interactions 
from Fermi's golden rule.
A similar treatment can be generalized to other contributions
to the collision integral as, e.g., electron-electron and impurity scattering or
coupling to external fields.  
 The Hamiltonian for an anharmonic crystal in presence of
electron-phonon and phonon-phonon interactions can be expressed as:
\begin{align}\label{eq:H}
\hat H = \hat H_{\rm e} +  \hat H_{\rm p} +  \hat H_{\rm ep} + \hat H_{\rm pp}\quad.
\end{align}
Here $\hat H_{\rm e}  = 
 \sum_{n\bk} \ve_{n\bk} \hat{c}^\dagger_{n\bk} \hat{c}_{n\bk}$
is the electronic Hamiltonian, 
where 
$ \hat{c}^\dagger_{n\bk}$ and $\hat{c}_{n\bk}$ are fermionic creation and annihilation operators, respectively. 
$\hat H_{\rm p} =  \sum_{\bq\nu }  \hbar\omega_{\bq \nu}[\hat a_{\bq \nu}  \hat a^\dagger_{\bq \nu} + 1/2]$
is the Hamiltonian of the lattice in the harmonic approximation. 
$\adq$ and $\aq$ are phonon creation and annihilation operators.
The eigenstates of $\hat H_{\rm p}$
can be expressed as  $\ket{ \chi_s } = \prod_{\bq\nu } \ket{  n^s_{\bq \nu}}$, where $ \ket{  n^s_{\bq \nu}}$ are the eigenstates of the number operator  $\hat N_{\bq\nu} = \adq\aq$
with eigenvalue $n^s_{\bq \nu}$. The quantity  $n^s_{\bq \nu}$ is the occupation number of the bosonic mode characterized by quantum numbers $\nu$ and $\bq$.
The superscript $s$-th indicates that the occupations are relative
to the $s$-th eigenstate $\ket{\chi_s}$ of the harmonic Hamiltonian.
With this notation, the energy of the harmonic lattice can be expressed as 
$ 
 E_s = \bra{\chi_s} \hat H_{\rm p} \ket{\chi_s} = \sum_{\bq\nu }
\hbar \omega_{\bq\nu} [  n^s_{\bq \nu} + 1/2]
$.

The third term in Eq.~\eqref{eq:H} is the electron-phonon coupling Hamiltonian:
\begin{align}\label{eq:Hep}
\hat H_{\rm ep} = N_{p}^{-\frac{1}{2}} \sum_{\substack{nm \nu\\ \bk\bq}} g^\nu_{mn}(\bk,\bq) \hat c_{m\bkq}^\dagger  \hat c_{n\bk} [ \hat a_{\bq \nu} + \hat a^{\dagger}_{-\bq \nu}]\quad,
\end{align}
where $N_{p}$ is the number of unit cells in the Born–von K\'arm\'an (BvK) supercell \cite{GiustinoRMP}.
The electron-phonon coupling matrix elements $g^\nu_{mn}(\bk,\bq)$
can be derived from first principles within the framework of
density-functional perturbation theory and they are defined as 
$g^\nu_{mn}(\bk,\bq) = \bra{\psi_{m\bkq}} \Delta_{\bq\nu} \hat v^{\rm KS} \ket{\psi_{n\bk}}$, 
where
$ \Delta_{\bq\nu} \hat v^{\rm KS}$ is the linear change
of the effective Kohn-Sham potential $ v^{\rm KS}$ due to a phonon
perturbation, and $\ket{\psi_{n\bk}}$ are single-particular orbitals,
solutions of the single-particle Kohn-Sham equations \cite{baroni01,GiustinoRMP}.

Finally, the last term in Eq.~\eqref{eq:H} is the
phonon-phonon coupling Hamiltonian, which
arises from anharmonicities of the lattice
and it can be expressed as \cite{Ziman1960}:
\begin{align}\label{eq:Hpp}
\hat H_{\rm pp} = \frac{1}{3!} &
\sum_{\bq\bq^{\prime}\bq^{\prime\prime}}
\sum_{\nu\nu^{\prime}\nu^{\prime\prime}}
\Psi_{\substack{\nu\nu^{\prime}\nu^{\prime\prime} \\ \bq\bq^{\prime}\bq^{\prime\prime}}}
[ \hat a_{\bq   \nu  } + \hat a^{\dagger}_{-\bq   \nu}]
[ \hat a_{\bq^{\prime}  \nu^{\prime} } + \hat a^{\dagger}_{-\bq^{\prime}  \nu^{\prime}}]
[ \hat a_{\bq^{\prime\prime} \nu^{\prime\prime}} + \hat a^{\dagger}_{-\bq^{\prime\prime} \nu^{\prime\prime}}]
\quad.
\end{align}
$\Psi_{\substack{\nu\nu^{\prime}\nu^{\prime\prime} \\ \bq\bq^{\prime}\bq^{\prime\prime}}}$
denotes the phonon-phonon scattering matrix elements, which is related to the
probability amplitude of three-phonon scattering processes.

\begin{figure}[t]
\begin{center}
   \includegraphics[width=0.5\textwidth]{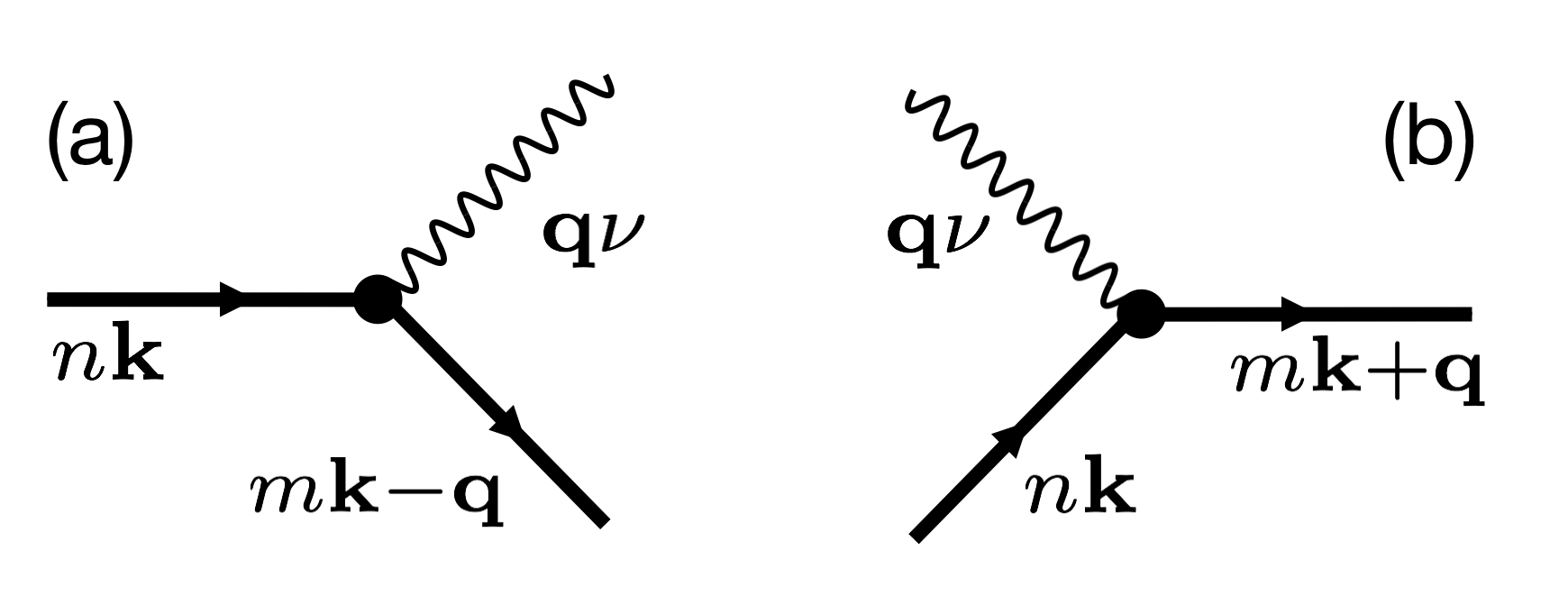}
\caption{\label{fig:eph}
Diagrammatic representation of (a) phonon-emission and (b) phonon-absorption processes.
Wavy lines represent non-interacting phonon propagator,
whereas straight lines denote non-interacting single-particle propagators.
}
\end{center}
\end{figure}

The derivation of the collision integrals for the Hamiltonian in Eq.~\eqref{eq:H} begins with the observation that electron-phonon and phonon-phonon interactions in solids are typically weak and, thus, the terms  $\hat{H}_{\rm ep}$ and $\hat{H}_{\rm pp}$ in the Hamiltonian in Eq.~\eqref{eq:H} can be treated as perturbations. 
Correspondingly, the rate $\Gamma_{i\rightarrow f}$ of transitions from an initial state  $ \ket{i} $ to a final
state $\ket{f}$ can be obtained via 
Fermi's golden rule:
\begin{align}\label{eq:FGR}
\Gamma_{i\rightarrow f} &= \frac{2\pi}{\hbar} | \bra{f} \hat V \ket{i} |^2   \delta(E^{\rm tot}_f - E^{\rm tot}_i)\quad,
\end{align}
where $E^{\rm tot}_i$ and $E^{\rm tot}_f$ are the total energies of the initial and
final states, respectively, and $\hat{V}$ is an arbitrary perturbation.
To proceed further, we focus on the electron-phonon interaction and we consider initial and final states in the form of a Born-Oppenheimer {\it ansatz} as $\ket{i} = \ket{\Psi_i}\ket{\chi_i} $, where 
$\ket{\chi_i}$  and $\ket{\Psi_i}$ are  eigenstates of the harmonic Hamiltonian $\hat H_{\rm p}$ and of the electronic Hamiltonian $\hat H_{\rm e}$, respectively. 
The matrix elements of the electron-phonon coupling Hamiltonian can be expressed as:
\begin{align}\label{eq:aux1}
\bra{f} \hat H_{\rm ep} \ket{i} =&
 N_{p}^{-\frac{1}{2}}
\sum_{{nm \nu }}
\sum_{{\bk\bq}}
g^{\nu}_{mn}(\bk,\bq) 
\bra{ \Psi_{f}} \hat c^\dagger_{m\bkq}  \hat c_{n\bk} \ket{ \Psi_i}
\bra{ \chi_{i}} \hat a_{\bq \nu} + \hat a^{\dagger}_{-\bq \nu}  \ket{ \chi_{i}}\quad.
\end{align}
The matrix elements of fermionic operators
in Eq.~\eqref{eq:aux1} differs from zero only if 
$\ket{ \Psi_i}$ and $\ket{ \Psi_f}$ differ only 
in the occupation of the ${n\bk}$ and ${m\bkq}$ states. In such case, it yields:
\begin{align} \label{eq:aux1_2}
\bra{ \Psi_{f}} \hat c^\dagger_{m\bkq}  
\hat c_{n\bk} \ket{ \Psi_i}  = [{f_{n\bk} (1-f_{m\bkq})}]^{\frac{1}{2}}\quad.
\end{align}
Similarly, from elementary considerations on the action of bosonic operators on the vibrational eigenstates $\ket{ \chi_{s}}$,
one can deduce that the matrix element $\bra{ \chi_{f}} \hat a_{\bq \nu} \ket{ \chi_{i}} = \sqrt{n_{\bq\nu}}$
(and similarly, $\bra{ \chi_{f}} \hat a^\dagger_{-\bq \nu} \ket{ \chi_{i}} = \sqrt{n_{-\bq\nu}+1}$) if the
state $\ket{ \chi_{f}}$ differs from $\ket{ \chi_{i}}$ only in the occupation of the
phonon $\bq\nu$ ($-\bq\nu$) such that  $n_{\bq\nu}^f =  n_{\bq\nu}^i -1 $ (and $n_{\bq\nu}^f =  n_{\bq\nu}^i +1 $), and it
vanishes otherwise.
The total rate (number of events per unit time) 
for the absorption of a phonon with quantum numbers $\bq\nu$ 
can be directly derived from  Eq.~\eqref{eq:FGR}
making use of Eqs.~\eqref{eq:aux1} and \eqref{eq:aux1_2}
and summing over all initial and final electronic states:
\begin{align}\label{eq:abs}
\Gamma^{\rm abs}_{\bq\nu} &= \frac{4\pi}{\hbar N_p} \sum_{mn\bk} 
| g^{\nu}_{mn}(\bk,\bq) |^2 
f_{n\bk} (1-f_{m\bkq})\delta(\ve_{n\bk} + \hbar\omega_{\bq\nu} - \ve_{m\bkq}) n_{\bq\nu}\quad.  
\end{align}
The energy difference between the initial and final states has been expressed as
$E^{\rm tot}_i - E^{\rm tot}_f = \ve_{n\bk} +\hbar\omega_{\bq\nu}- \ve_{m\bkq} $.
The diagrammatic representation of a phonon absorption process of this kind is 
reported in Fig.~\ref{fig:eph}~(b). 
The same procedure can be repeated by considering phonon emission processes [Fig.~\ref{fig:eph}~(a)], yielding:
\begin{equation}\label{eq:em_2}
\Gamma^{\rm em}_{\bq\nu} = \frac{4\pi}{\hbar N_p} \sum_{mn\bk} 
| g^{\nu}_{mn}(\bk,\bq) |^2 
f_{n\bk} (1-f_{m\bkq})\delta(\ve_{n\bk} - \hbar\omega_{\bq\nu} - \ve_{m\bkq}) (n_{\bq\nu}+1) \quad.
\end{equation}

The total rate of change in the phonon occupation $n_{\bq\nu}$ 
due to the electron-phonon interactions  $\Gamma^{\rm ep}_{\bq\nu}$ can thus be defined as the difference between
the rates of phonon emission ($\Gamma^{\rm em}_{\bq\nu}$) and absorption ($\Gamma^{\rm abs}_{\bq\nu}$) processes:
\begin{align}\label{eq:Gamma_pe}
\Gamma^{\rm pe}_{\bq\nu}(t) =& \frac{4\pi}{\hbar N_p} \sum_{mn\bk} 
| g^{\nu}_{mn}(\bk,\bq) |^2 f_{n\bk} (1-f_{m\bkq})\\
 &\times [ \delta(\ve_{n\bk} - \hbar\omega_{\bq\nu} - \ve_{m\bkq}) (n_{\bq\nu}+1)  
 - \delta(\ve_{n\bk} + \hbar\omega_{\bq\nu} - \ve_{m\bkq}) n_{\bq\nu} ] \quad. \nonumber
\end{align}
Equation~\eqref{eq:Gamma_pe} is the
phonon collision integral due to the electron-phonon interaction. 
Its time dependence arises from the changes of the electron and phonon distribution functions ($f_{n\bk}$ and $n_{\bq\nu}$) over time. 
In conditions of thermal equilibrium between electrons and the lattice,
as for instance in an ideal situation in which electron and phonon
occupations are described by Fermi-Dirac and Bose-Einstein statistics 
at a given temperature, the rates $\Gamma^{\rm abs}_{\bq\nu}$
and $\Gamma^{\rm em}_{\bq\nu}$ are equal and opposite in sign, 
indicating that the total change in the phonon number $n_{\bq\nu}$ vanishes,
since the emission and absorption of phonons are perfectly compensated.

\begin{figure}[t]
\begin{center}
   \includegraphics[width=0.98\textwidth]{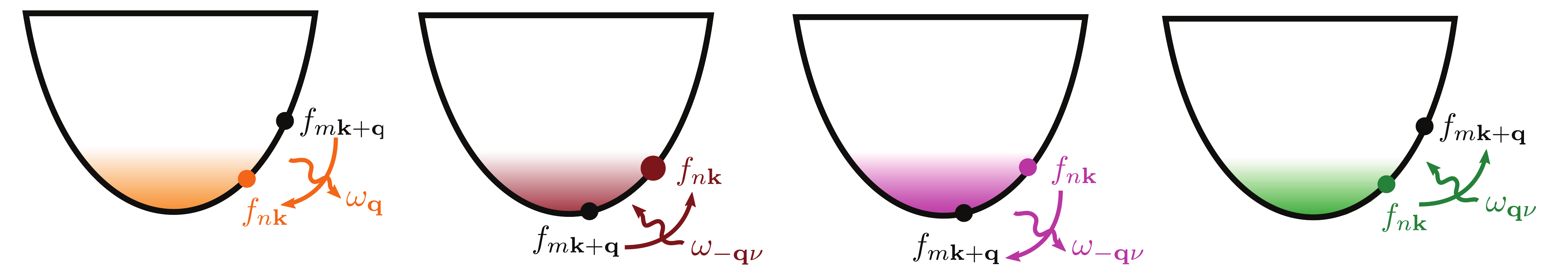}
\caption{\label{fig:pon}
Schematic representation of the four phonon-assisted scattering processes
included in the electron collision integral due to the electron-phonon
interaction (Eq.~\eqref{eq:Gamma_el}).  Adapted from Ref.~\cite{Ponce2020}.
}
\end{center}
\end{figure}

Following similar steps, the electronic collision integral due to the electron-phonon interaction can be derived as:
  \begin{align} \label{eq:Gamma_el}
   \Gamma^{\rm ep}_{{n\bk}}(t) &=  \frac{2\pi}{\hbar N_p}
  \sum_{m\nu\bq}  \, |g_{mn}^\nu(\bk,\bq)|^2  \\
\times & \big\{(1-f_{n\bk})f_{m\bk+\bq}   \d(\ve_{n\bk}-\ve_{m\bk+\bq} + \hbar\w_{\bq\nu})(1+ n_{\bq\nu})   \nonumber\\&     +(1-f_{n\bk})f_{m\bk+\bq}    \d(\ve_{n\bk}-\ve_{m\bk+\bq} - \hbar\w_{\bq\nu})n_{\bq\nu}  \nonumber  \\
 &      -f_{n\bk}(1-f_{m\bk+\bq})   \d(\ve_{n\bk}-\ve_{m\bk+\bq} - \hbar\w_{\bq\nu})(1+ n_{\bq\nu})  \nonumber\\
 &      -f_{n\bk}(1-f_{m\bk+\bq})   \d(\ve_{n\bk}-\ve_{m\bk+\bq} + \hbar\w_{\bq\nu})n_{\bq\nu} \big\}\quad.\nonumber
  \end{align}
Each term in this expression is directly related to phonon-assisted 
electronic transitions involving states in the vicinity of the Fermi surface, as depicted in Fig.~\ref{fig:pon}. 
In particular, the first and second terms in Eq.~\eqref{eq:Gamma_el} arise
from processes in which an electron scatter from $m\bkq$ in to $n\bk$
via the emission and absorption of a phonon, respectively.
The third and fourth terms arise from the scattering from  $n\bk$ into  $m\bkq$
due to phonon-emission and absorption processes.
In analogy to Eq.~\eqref{eq:Gamma_pe}, at thermal equilibrium,
scattering processes in and out of $n\bk$ balance each other, leading to $   \Gamma^{\rm ep}_{{n\bk}} =0$. Note that the rate $\Gamma^{\rm ep}_{{n\bk}}$ of electron population relaxation defined with Eq. \eqref{eq:Gamma_el} and the corresponding lifetime $\tau^{\rm ep}_{{n\bk}} =\hbar / \Gamma^{\rm ep}_{{n\bk}}$ should not be confused with the quasiparticle decay rate and lifetime as obtained from the imaginary part of the electron self-energy \cite{yang15}. It is the former and not the latter that is usually extracted from pump-probe measurements.

The derivation of the scattering rate due to the phonon-phonon scattering
involves the tedious (but otherwise straightforward) evaluation of several
matrix elements of bosonic operators. The result can be recast in the form: 
\begin{align}
\Gamma^{\rm pp}_{{\bq\nu}}(t) &= \frac{ 2\pi }{\hbar}\label{eq:Gamma_pp}
\sum_{\nu'\nu''} \!\int\!\!\frac{d\bq'}{\Omega_{\rm BZ}} \left|\Psi_{\mathbf{q
q'}^{\prime} \mathbf{q}^{\prime \prime}}^{\nu \nu^{\prime}\nu''}\right|^{2}
\\ \nonumber  \times 
&\left[\right.  [(n_{\mathbf{q} \nu}+1)(n_{\mathbf{q}^{\prime}
\nu^{\prime}}+1) n_{\mathbf{q}^{\prime \prime} \nu^{\prime
\prime}}-n_{\mathbf{q} \nu} n_{\mathbf{q}^{\prime}
\nu^{\prime}}(n_{\mathbf{q}^{\prime \prime} \nu^{\prime \prime}}+1)]
\delta\left(\omega_{\mathbf{q} \nu}+\omega_{\mathbf{q}^{\prime}
\nu^{\prime}}-\omega_{\mathbf{q}^{\prime \prime} \nu^{\prime
\prime}}\right)\delta^{\bf G}_{\bq\bq'-\bq''}+\\
\frac{1}{2}&\left.\left[\left(n_{\mathbf{q} \nu}+1\right)
n_{\mathbf{q}^{\prime} \nu^{\prime}} n_{\mathbf{q}^{\prime \prime} \nu^{\prime
\prime}}-n_{\mathbf{q} \nu}\left(n_{\mathbf{q}^{\prime}
\nu^{\prime}}+1\right)\left(n_{\mathbf{q}^{\prime \prime} \nu^{\prime
\prime}}+1\right)\right] \delta\left(\omega_{\mathbf{q}
\nu}-\omega_{\mathbf{q}^{\prime} \nu^{\prime}}-\omega_{\mathbf{q}^{\prime
\prime} \nu^{\prime \prime}}\right) \delta^{\bf G}_{\bq-\bq'-\bq''}\right]
\nonumber
\end{align}
where the modified Kronecker's $\delta^{\bf G}_{\bq}$ equals unity if $\bq =0 $
or $\bq ={\bf G} $, where ${\bf G}$ is a reciprocal-lattice vector, and it is zero otherwise.
Equations~\eqref{eq:Gamma_pp} vanishes identically if the lattice is at thermal
equilibrium. 

Combining Eqs.~\eqref{eq:Gamma_el}-\eqref{eq:Gamma_pp}, the time-dependent
Boltzmann equation can be rewritten as: 
\begin{align}
\label{eq:bte_f}  {\D_t f_{n\bk}(t)}     &= \Gamma^{\rm ep}_{n{\bk}}[f_{n\bk}(t), n_{\bq\nu}(t)]\quad, \\
\label{eq:bte_n}  {\D_t n_{\bq \nu}(t)}  &= \Gamma^{\rm pe}_{{\bq\nu}}[f_{n\bk}(t), n_{\bq\nu}(t)]  + \Gamma^{\rm pp}_{{\bq\nu}}[n_{\bq\nu}(t)]\quad.
\end{align}
A numerical procedure to solve Eqs.~\eqref{eq:bte_f} and \eqref{eq:bte_n} consists in: 
(i) defining an initial electronic (or vibrational) excited state 
characterized by electronic (vibrational) occupations which differs from the 
equilibrium ones (as e.g. in Fig.~\ref{fig:car21}); (ii) solve the differential 
equation using iterative methods (as, e.g., the Euler of Runge-Kutta methods); 
(iii) update the collision integrals via Eqs.~\eqref{eq:Gamma_el}-\eqref{eq:Gamma_pp} 
at each time step. 

Besides the aforesaid microscopic scattering process, electron-electron interaction plays a major role in thermalization of electron-lattice system, especially in the 10-fs timescale. The corresponding collision integral $\Gamma^{\rm ee}_{n{\bk}}[f_{n\bk}(t)]$ enters Eq.\,\eqref{eq:bte_f} and it is usually defined in terms of electron scatterings $\{\mathbf{k},\mathbf{k}'\}\rightarrow \{\mathbf{k}+\mathbf{q},\mathbf{k}'-\mathbf{q}\}$ (and vice versa) mediated by the statically screened Coulomb interaction. For a detailed description of electron-electron scattering processes we refer the reader to Refs.\,\cite{rana07,baranov14,ono18,Ono2020}.
Note that the electron-plasmon scatterings, as a part of the dynamical electron-electron interaction, is as well considered to be essential in the early stage of electron thermalization process \cite{pines62,hamm16,kim21}, however, these dynamical effects are rearly taken into account when studying hot carrier thermalization.

\subsection{Relation between the TTM and the time-dependent Boltzmann equation}\label{sec:bte-ttm} 

As demonstrated by Allen~\cite{Allen1987}, the thermalization dynamics of electrons and phonons in presence of the electron-phonon interaction can be recast in the form of a TTM, as formulated via Eqs.~\eqref{eq:mod0a} and \eqref{eq:mod0b}, starting entirely from first principles. 
In short, by expressing the coupling parameter $g$ in terms of the Eliashberg function $\alpha^2F$, all free parameters of the models are fixed. A detailed derivation of this scheme can be found, for instance, in Refs.~\cite{maldonado_theory_2017,ono18}.

In the following, we report an alternative derivation of the 
TTM with scope of emphasizing the link with the phonon lifetime, as obtained from phonon self-energy 
due to the electron-phonon interaction. 
We begin by considering the total energy of a set of non-interacting electrons ($E_{\rm el}$) 
and phonons ($E_{\rm ph}$): 
\begin{align}
E_{\rm el}(t) &= N_{p}^{-1} \sum_{n}\sum_{\bk} \ve_{n\bk} f_{n\bk}(t) \quad, \\
E_{\rm ph}(t) &= N_{p}^{-1} \sum_{\nu}\sum_{\bq} \hbar\omega_{\bq\nu} \left[  n_{\bq\nu}(t)  + \frac{1}{2} \right ] \quad.
\end{align}
The rate of change of the energies can be expressed as: 
\begin{align}\label{eq:cel1}
C_{\rm el} \D_t T_{\rm el} &=  N_{p}^{-1} \sum_{n}\sum_{\bk}  \ve_{n\bk}  \D_t f_{n\bk}\quad.\\
\label{eq:cel2}
C_{\rm ph} \D_t T_{\rm ph} &= N_{p}^{-1} \sum_{\nu}\sum_{\bq} \hbar \omega_{\bq\nu}  \D_t n_{\bq\nu} \quad,
\end{align}
where ${\D_t} = \D / \D t$, 
the left-hand side has been  rewritten making use of the chain rule $\D_t E = \frac{\D E}{\D T} \frac{\D T}{\D t}$, 
and the heat capacity $C_{\rm el(ph)} = {\D E_{\rm el(ph)}}/{\D T_{\rm el(ph)}}$ has been introduced. 
Equations~\eqref{eq:cel1} and \eqref{eq:cel2} rely on the assumption that electron energies $\ve_{n\bk}$ and phonon 
frequencies $\omega_{\bq\nu}$ do not depend on time.
The time derivative of the fermionic and bosonic distribution functions entering Eqs.~\eqref{eq:cel1} and \eqref{eq:cel2}
can be obtained via the time-dependent Boltzmann equation.

A central assumption of the non-thermal lattice model is 
that, at each time $t$, the electronic occupation $f_{n\bk}$ 
can be approximated by a Fermi-Dirac function at temperature $T_{\rm el}$:
\begin{align}
f_{n\bk} \simeq \{\exp[( \ve_{n\bk}-\mu) /k_{\rm B}T_{\rm el}] +1 \}^{-1}\quad.
\end{align}
Similarly, it is assumed that the lattice remains at thermal equilibrium throughout the dynamics, that is, 
all vibrational modes are at the same temperature $T_{\rm ph}$ (a condition that 
may be violated in the non-equilibrium dynamics of the lattice following photo-excitation \cite{caruso/2021}). 
Under these assumptions, the arguments of the collision integrals can be simplified. 
For instance, the argument of Eq.~\eqref{eq:Gamma_pe} can be rewritten as \cite{maldonado_theory_2017}: 
\begin{align}\label{eq:int2}
&[\left(1-f_{n \bk}\right)f_{{m} \bk+\bq}\left(n_{\bq \nu}+1\right) 
       -f_{n \bk}\left(1-f_{m \bk+\bq}\right)  n_{\bq \nu}] \\ 
& = ( f_{{m} \bk+\bq}  - f_{n \bk} ) [ n_{\rm BE}(\omega_{\bq\nu},T_{\bq\nu}) - n_{\rm BE}(\omega_{\bq\nu}, T_{\rm el})] \nonumber 
\quad,
\end{align}
where we introduced the Bose-Einstein function: 
\begin{align}
n_{\rm BE}(\omega,T)  = [ \exp({{\hbar\omega}/{k_{\rm B}T}}) -1 ]^{-1}\quad.
\end{align}
Additionally, we set $\ve_{{m} \bk+\bq} - \ve_{{n} \bk}= \hbar\omega_{\bq\nu}$ because of the Dirac-$\delta$ in Eq.~\eqref{eq:Gamma_pe}, and we used $n_{\bq\nu} = n_{\rm BE}(\omega_{\bq\nu},T_{\rm ph})$. 
A Tailor expansion of $n_{\rm BE}$ to first order yields: 
\begin{align*} 
 n_{\rm BE}(\omega_{\bq\nu},T_{\rm ph}) &- n_{\rm BE}(\omega_{\bq\nu}, T_{\rm el}) 
\simeq    (T_{\rm ph}-T_{\rm el} ) \left.\frac{\partial n_{\rm BE(\omega_{\bq\nu},T)} } {\partial T}\right|_{{T=T_{\rm ph}}} 
=  (T_{\rm ph}-T_{\rm el} ) \frac{c_{\bq\nu}}{\hbar\omega_{\bq\nu}}\quad.
\end{align*}
In the last equality, we introduced the specific heat of a single harmonic oscillator 
$c_{\bq\nu} = \hbar\omega_{\bq\nu} {\partial n_{\bq\nu}}/ {\partial T}$. 

Making use of Eqs.~\eqref{eq:int2}, the integrand within the phonon-electron collision integral in  
Eq.~\eqref{eq:Gamma_pe} can be rewritten as: 
\begin{align}\label{eq:Gamma_ph1}
 \Gamma^{\rm pe}_{{\bq\nu}} 
& = \frac{c_{\bq\nu}}{\hbar\omega_{\bq\nu}}\frac{ (T_{\rm el}- T_{\rm ph} )} {  \tau_{\bq\nu}}\quad.
\end{align}
Where we introduced the phonon lifetime due to the electron-phonon interaction \cite{GiustinoRMP}: 
\begin{align}\label{eq:T3}
 \tau^{-1}_{\bq\nu}&= \frac{ 4\pi}{\hbar N_p} \sum_{m n \bk} |g_{mn\nu}(\bk,\bq)|^2  (  f_{n \bk}- f_{{m} \bk+\bq}   )
 \delta\left(\ve_{m \bk+\bq}-\ve_{n \bk}-\hbar \omega_{\bq \nu}\right)\quad.
\end{align}
Via the identity $\tau_{\bq\nu}^{-1}=-2\,{\rm Im}\,\Pi_{\bq\nu} = $,
Eq.~\eqref{eq:Gamma_ph1} further establishes a simple relation between the phonon self-energy due to the 
electron-phonon interaction $\Pi_{\bq\nu}$ and the corresponding collision integral. 
The limit of validity of this identity are defined by the assumptions made thus far. 
Combining \eqref{eq:T3} with Eq.~\eqref{eq:cel2}: 
\begin{align}
C_{\rm ph} \D_t T_{\rm ph} 
&= N_{p}^{-1} \sum_{\nu}\sum_{\bq} c_{\bq\nu} \frac{ (T_{\rm el}- T_{\rm ph} )} {  \tau_{\bq\nu}}  
\quad.
\end{align}
A similar identity can be derived by applying the same procedure for the electronic term.  
After introducing the effective electron-phonon coupling constant $g$: \begin{align}\label{eq:g}
g = N_{p}^{-1} \sum_{\nu}\sum_{\bq} 
\frac{c_{\bq\nu}} { \tau_{\bq\nu}}  \quad,
\end{align}
the equations of the TTM, Eq.~\eqref{eq:mod0a} and \eqref{eq:mod0b} are recovered.

\section{Ultrafast carrier dynamics in graphene}\label{sec:graphene}

\begin{figure}[t]
\centering
\resizebox*{14cm}{!}{\includegraphics{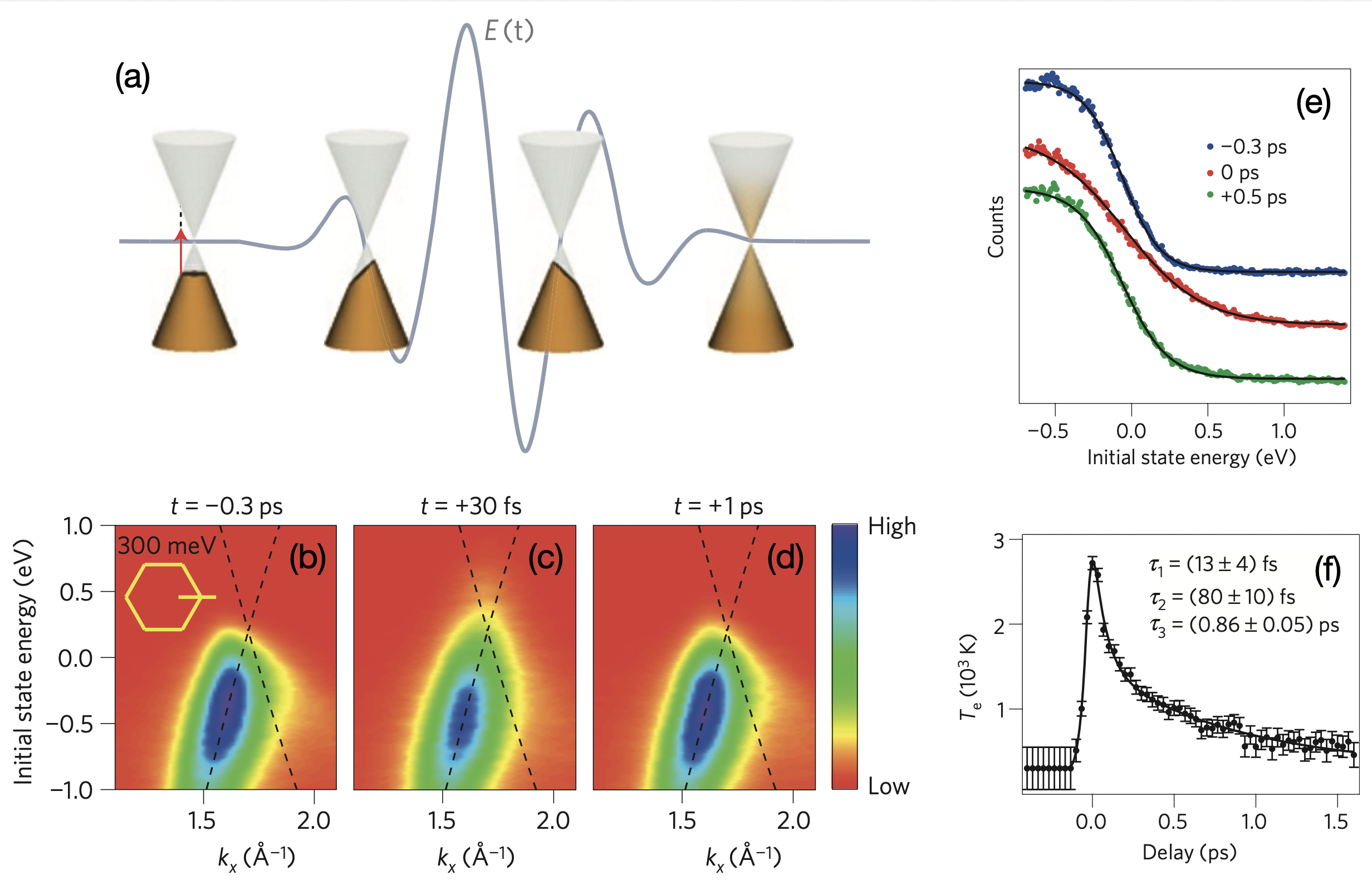}}
\caption{
(a) Schematic representation of the thermalization dynamics of electrons and
holes following photo-excitation in the vicinity of the Dirac cone of graphene.
Time and angle-resolved photoemission spectral function of doped graphene
before photo-excitation (b), and at time delays $t=30$~fs (c), and $t=1$~ps (d).
Momentum integrated spectral function (e) and, superimposed as a continuous
line, the Fermi-Dirac function at the effective electronic temperature $T_e$.
(f) Time-dependent effective electronic temperature derived by the fitting
procedure illustrated in (e).  Adapted from Ref.~\cite{gierz_snapshots_2013}.
} 
\label{fig:gie}
\end{figure}

Graphene constitutes a paradigmatic example to introduce the hot-carrier 
dynamics revealed by pump-probe experiments in 2D materials.
Ultrafast phenomena in graphene have been extensively investigated 
experimentally to explore its suitability for optoelectronic 
applications and light harvesting. Additionally, a recent discovery  
of the light-induced Hall effect has revealed a novel route  
to trigger a non trivial topological behaviour using light pulses 
\cite{mciver_light-induced_2020,PhysRevX.10.041013,sentef_theory_2015}. 
The first ultrafast experimental studies resorted to optical techniques
\cite{kampfrath_strongly_2005,george_ultrafast_2008,sun_ultrafast_2008,breusing_ultrafast_2009,
wang_ultrafast_2010,lui_ultrafast_2010,breusing_ultrafast_2011,winnerl_carrier_2011,li_femtosecond_2012,jnawali13,frenzel14,jensen14}.
Subsequently, time- and angle-resolved photoemission 
spectroscopy (tr-ARPES) enabled to directly monitor 
the non-equilibrium dynamics of photo-excited carriers
with momentum resolution \cite{ishida11,johannsen_direct_2013,gierz_snapshots_2013,ulstrup14,gierz14_faraday,gierz_tracking_2015,gierz_phonon-pump_2015,johannsen15,aeschlimann_ultrafast_2017,yang17,pomarico17,someya17}. Despite over a decade of extensive research, ultrafast hot carrier dynamics in graphene is still actively explored \cite{tan17,tomadin18,na19,caruso_photoemission_2020,novko2019,degiovannini20,novko_first-principles_2021,zhang21,degiovannini22}.

\subsection{Experimental signature of hot-carrier dynamics in graphene}\label{sec:Ca}

The ultrafast dynamics of Dirac carriers in pump-probe
experiments is usually approximated 
as a four-step process consisting of (i) photo-excitation, (ii) formation of a quasi-equilibrium state, (iii)
full electron thermalization, and (iv) energy transfer to the lattice (acoustic phonons).
In the step (i), the interaction with a pump pulse 
drives the system into an electronic excited state, 
with carriers photo-excited above the Fermi level. With this a non-equilibrium electron distribution is created. The step (ii) involves electron-electron scatterings, impact ionization, and Auger processes (timescale of $\sim 10$\,fs) \cite{rana07,GIERZ201753}, which promote photo-excited electrons and holes towards the Fermi level.
Some experiments indicate that under suitable conditions 
a regime of population inversion can be established, i.e., two electron distributions with separate chemical potentials (and temperatures)
\cite{gierz_snapshots_2013,gierz_population_2015}. 
Interestingly, some theoretical works suggested that strong scatterings between non-equilibrium electrons and phonon are active already at this stage \cite{baranov14,hu22}. Even more, this early electron-phonon scattering process might be responsible for majority of energy flow, leaving a small portion of excess energy for the later scatterings between equilibrium electrons and phonons \cite{baranov14}.
In step (iii), Auger recombination \cite{winzer_carrier_2010,winnerl_carrier_2011}, scatterings with optical phonons \cite{kampfrath_strongly_2005,johannsen_direct_2013}, as well as plasmon emission \cite{hamm16,kim21}
bring the electrons to a thermalized regime ($\sim100$\,fs), where the 
electronic distribution function is described by a 
high-temperature Fermi-Dirac function. 
Finally, in step (iv), the photo-excited 
electrons and holes dissipate their energy via phonon-assisted 
scattering processes, mostly via the acoustic modes ($\sim 1-10$\,ps).
Supercollisions \cite{johannsen_direct_2013} and the formation of 
hot phonons \cite{doi:10.1021/nl301997r} have been proposed as 
underlying physical processes to explain the energy transfer to 
the lattice \cite{kampfrath_strongly_2005,pogna21}. Further experimental investigations on the parent compound graphite 
corroborated the hot-phonon picture \cite{stange_hot_2015,rohde_ultrafast_2018}. 
Overall, these studies revealed a complex non-equilibrium dynamics 
characterized by the coexistence of several scattering mechanisms 
which are challenging to decipher based on purely experimental investigations.  

Exemplary tr-ARPES experiments on hole-doped graphene are illustrated in 
Fig.~\ref{fig:gie} \cite{gierz_snapshots_2013}. 
As a result of substrate-induced hole-doping, the Fermi energy is 
located 0.2~eV below the Dirac point [Fig.~\ref{fig:gie}~(a)]. 
Figures~\ref{fig:gie}~(b-d) illustrate several time snapshots of the 
tr-ARPES spectral function in the vicinity of the Dirac point.  
Before excitation [Fig.~\ref{fig:gie}~(b)], the spectral intensity reflects occupied electronic states
at equilibrium, with broadening arising from finite temperature and 
experimental resolution. Owing to the optical selection rules governing 
the coupling with linearly-polarized light, the spectral intensity is dominated 
by contributions arising from the left sub-band \cite{PUSCHNIG2015193}. 
The change in spectral intensity for $t=30$~fs [Fig.~\ref{fig:gie}~(c)] 
reflects the excitation of carriers above the Fermi level, 
whereas measurements at longer time delays [Fig.~\ref{fig:gie}~(d)] 
indicates the recovery of an equilibrium regime.

\begin{figure}[t]
\centering
\resizebox*{14cm}{!}{\includegraphics{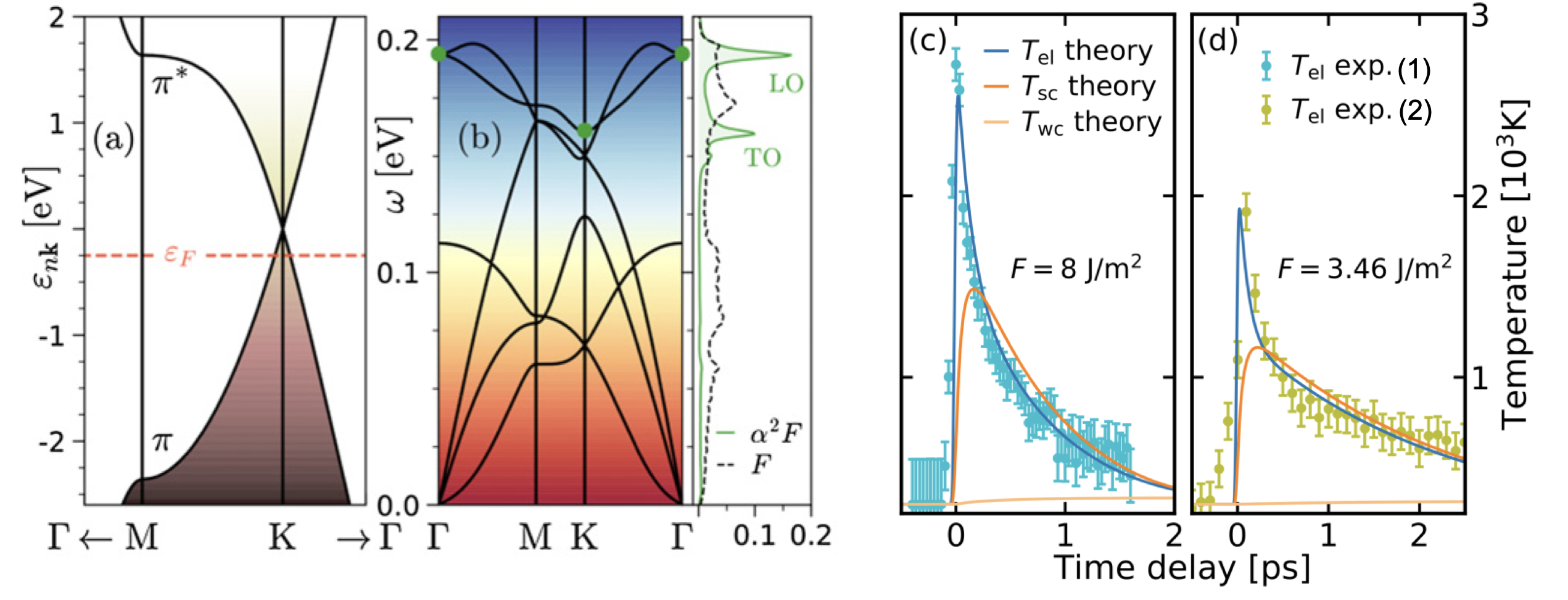}}
\caption{(a) DFT Band structure of graphene in the vicinity of 
the Dirac cone along the M-K high-symmetry path in the Brillouin zone. 
(b) Left: Phonon dispersion of graphene obtained from DFPT. 
Right: Phonon density of states ($F$) and Eliashberg function ($\alpha^2F$) of graphene.
The peaks at 0.16 and 0.19~eV reflect the strong coupling to 
transverse optical (TO) and longitudinal optical (LO)
phonons in the vicinity of the K and $\Gamma$ points, respectively [green dots in panel (b)]. 
(c-d) Pump-probe photoemission measurements of the effective electronic temperature $T_{\rm el}$ 
of graphene (dots) for photo-excitation fluences $F=8~{\rm J/m}^2$ (c) 
and (d) $F=3.46~{\rm J/m}^2$ from Refs.~\cite{gierz_snapshots_2013,johannsen_direct_2013}. 
Simulations based on the NLM (more precisely, three-temperature model) with first-principles parameters are reported as continuous lines. 
Panels (a-b) and (c-d) are adapted from Ref.~\cite{novko_first-principles_2021} 
and Ref.~\cite{caruso_photoemission_2020}, respectively. 
} 
\label{fig:C}
\end{figure}

High-quality tr-ARPES measurements can be analysed to extract 
the effective electronic temperatures $T_{\rm el}$ and its time dependence
\cite{fann_direct_1992,fann_electron_1992,wang_time-resolved_1994,ishida11,johannsen_direct_2013,gierz_snapshots_2013,andreatta19,majchrzak21},
thus, establishing a direct link with the TTM and NLM discussed in Secs.~\ref{sec:model} and \ref{sec:NLM}.
For a given time delay,  $T_{\rm el}$ can be determined by fitting 
the normalized momentum-integrated photoemission intensity with a 
Fermi-Dirac function. 
This procedure is illustrated in Fig.~\ref{fig:gie}~(e) where 
the fit Fermi-Dirac function (continuous line) is superimposed to 
the measured energy distribution curves (dots). 
The time dependence of  $T_{\rm el}$, shown  in Fig.~\ref{fig:gie}~(f), 
is obtained by repeating this procedure for several measured time delays 
and it closely resembles the trend reported for the light-driven TTM  
illustrated in Fig.~\ref{fig:sketch1}~(c):  
the photo-excitations of the electrons by the pump pulse 
manifests itself through the initial increase of electronic temperature, 
whereas the subsequent cooling reflects the thermalization dynamics due 
electron-phonon scattering processes.

\subsection{Theoretical modelling of carrier thermalization in graphene }\label{sec:Cb}

First-principles calculations of the electron-phonon interaction can be combined  
with the time-propagation algorithms discussed in Secs.~\ref{sec:model}-\ref{sec:bte}  
to investigate the origin of the hot-carrier dynamics and its fingerprints in spectroscopy. 
The band structure and phonon dispersion of hole-doped graphene 
is illustrated in Fig.~\ref{fig:C}~(a) and (b), respectively \cite{novko_first-principles_2021}, 
whereas the phonon density of state ($F$) and the Eliashberg function ($\alpha^2 F$)
are shown in the right panel of Fig.~\ref{fig:C}~(b). 
The Eliashberg function reflects the weighted
phonon density of states, to which individual phonons 
contribute according to their electron-phonon coupling strength \cite{grimvall1981electron}. 
The sharp peaks in $\alpha^2F$ at 160 and 190~meV indicate 
that the electron-phonon coupling arises primarily from longitudinal 
optical (LO or $E_{2g}$) phonons at $\Gamma$ and transverse optical (TO or $A_{1}'$) modes at K 
[green dots in Fig.~\ref{fig:C}~(b)], whereas remaining phonons couple relatively 
weakly to the electrons. This finding suggests that a few 
phonons are likely to provide a preferential decay channel 
for the relaxation of excited carriers, leading to the emergence of hot phonons -- 
namely, phonons characterized by a higher vibrational temperature. 
The pronounced anisotropy of the electron-phonon interaction in graphene
can be explicitly accounted for by formulating a NLM (i.e., a three-temperature model) which discriminates 
between strongly- (sc) and weakly-coupled (wc) phonons \cite{caruso_photoemission_2020}, 
and by determining the model parameters from first-principles calculations \cite{baroni01,Giannozzi_2017,Ponce2016}.  
The effective electronic temperature $T_{\rm el}$ 
and the vibrational temperatures ($T_{\rm wc}$ and $T_{\rm sc}$) 
obtained from the solution of the NLM [Eqs.~\eqref{eq:NLMa} and \eqref{eq:NLMb}]
are illustrated in Figs.~\ref{fig:C}~(c) and (d) for different excitation 
conditions  \cite{caruso_photoemission_2020}. 
The electronic temperature follows the characteristic trend expected 
for photoexcited electrons and it is in good agreement with 
the experimental values extracted from tr-ARPES.  
The preferential excitation of strongly-coupled modes upon carrier relaxation
leads to a pronounced rise of $T_{\rm sc}$, whereas the vibrational 
temperatures of weakly-coupled modes $T_{\rm wc}$  is smaller.

\begin{figure}[t]
\centering
\resizebox*{14cm}{!}{\includegraphics{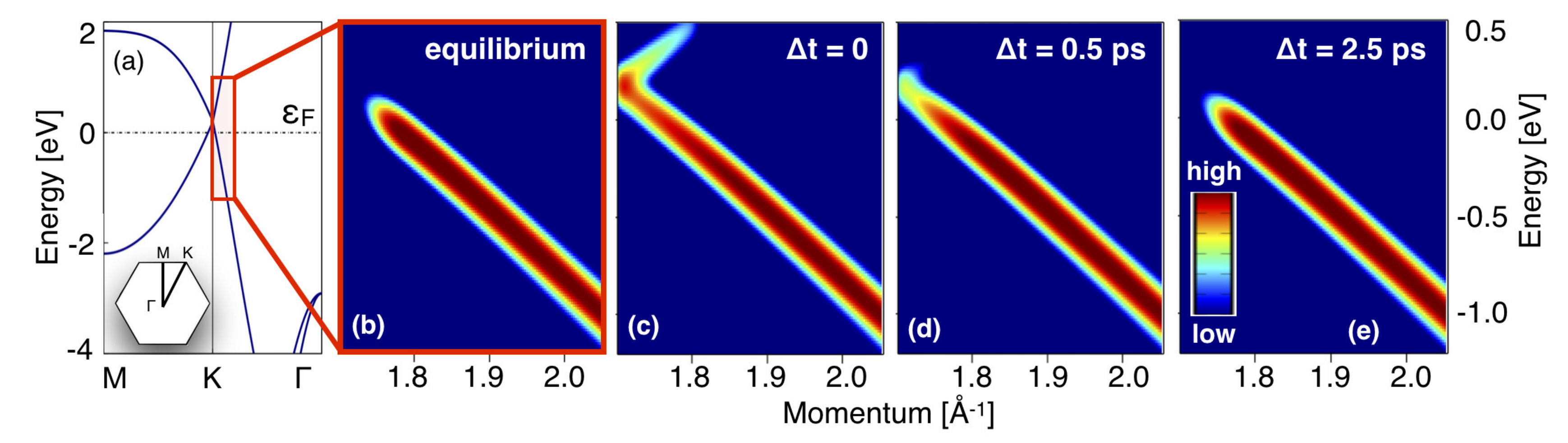}}
\caption{(a) DFT Band structure of graphene in the vicinity of the Dirac cone.
The dashed line denotes the Fermi energy $\ve_{\rm F} = -0.2$~eV, corresponding
to a carrier concentration $8\times10^{12}$~cm$^{-2}$. The inset illustrates
the hexagonal Brillouin zone.
First-principles calculations of the time- and angle-resolved spectral function
obtained by considering electronic and vibrational occupations derived from the
solution of the TTM before excitation (b), and at time delays $\Delta t =0$
(c), $\Delta t =0.5$~ps (d), and $\Delta t=2.5~$ps (e).   Reproduced from
Ref.~\cite{caruso_photoemission_2020}.} 
\label{fig:car}
\end{figure}

By post-processing the effective temperatures  $T_{\rm el}$, $T_{\rm wc}$, and $T_{\rm sc}$, 
one can define a simple procedure to estimate the transient changes 
of many-body interactions and their signatures in photoemission. 
The self-energy due to electron-phonon interaction can be 
modified to account for changes of the electronic and 
vibrational distribution functions as: 
  \begin{align}\label{eq:sigma}
    \Sigma_{n{\bf k}} (\omega,\Delta t) &= 
\frac{1}{N_p} \sum_{m\nu \bq}
       |g_{mn\nu}({\bf k},{\bf q})|^2  
\\  \times & \left[ \frac { n_{{\bf q}\nu}(\Delta t) + f_{m{\bf k+q}} (\Delta t) }
    {\hbar\omega - \tilde\varepsilon_{m{\bf k+q}} + \hbar\omega_{{\bf q}\nu} }
    + \frac { n_{{\bf q}\nu}(\Delta t) + 1 - f_{m{\bf k+q}} (\Delta t) }
    {\hbar\omega - \tilde\varepsilon_{m{\bf k+q}} - \hbar\omega_{{\bf q}\nu} } \right]\quad.\nonumber
  \end{align}
The time dependence arise from the dependence of  
the electron and phonon distribution functions on the 
effective temperatures, given by 
$f_{n{\bf k}}  = [e^{\varepsilon_{n{\bf k}} / k_{\rm B} T_{\rm el}} + 1 ]^{-1}$
and $n_{\nu{\bf q}} = [e^{\omega_{\nu{\bf q}}    / k_{\rm B} T_{\nu}} - 1 ]^{-1}$, respectively, 
where  $T_{\nu}= T_{\rm sc}$ ($T_{\nu}= T_{\rm wc}$) for the strongly (weakly) coupled modes.
The tr-ARPES spectral function can be directly derived from Eq.~\eqref{eq:sigma} via
$A_{\bf k}(\omega, \Delta t) =
{\pi^{-1}} \sum_n {\rm Im}\,[\hbar\omega - \varepsilon_{n{\bf k}}-  \Sigma_{n{\bf k}} (\omega, \Delta t)]^{-1}$. 
The spectral function of graphene obtained from this procedure is illustrated
in Fig.~\ref{fig:car}, and it reproduces the main spectral signatures of 
photoexcitations revealed by experiments as well as the characteristic 
time scales of photoexcitations.  

Other time-dependent physical features of ultrafast hot carrier cooling visible in pump-probe spectroscopy can be theoretically captured in a similar fashion, i.e., by benefiting from the time dependence of electron and phonon temperatures. For instance, time dynamics of electron excitations (e.g., plasmons and screened interband transitions) immediately after the laser excitation can be simulated via optical conductivity $\sigma(\omega;T_{\rm el},T_{\rm ph})$ or dielectric function $\epsilon(\mathbf{q},\omega;T_{\rm el},T_{\rm ph})$ \cite{sun_femtosecond-tunable_1994,dalconte12}. The spectral function of electron excitations (or the so-called electron energy loss function) is defined as \cite{novko_first-principles_2021,kupcic14}
\begin{align}
-\mathrm{Im}\,\left[ \frac{1}{\epsilon(\mathbf{q},\omega;T_{\rm el},T_{\rm ph})} \right]=-\mathrm{Im}\,\left[ \frac{\omega^2}{\omega^2-\Omega^2(\mathbf{q},\omega;T_{\rm el},T_{\rm ph})+i\Gamma(\mathbf{q},\omega;T_{\rm el},T_{\rm ph})} \right]\quad,
\label{eq:epsilon_t}
\end{align}
where $\Omega^2(\mathbf{q},\omega;T_{\rm el},T_{\rm ph})=v(\mathbf{q})q^2 {\rm Re}\,\pi_{\alpha\alpha}(\mathbf{q},\omega;T_{\rm el},T_{\rm ph})$ and $\Gamma(\mathbf{q},\omega;T_{\rm el},T_{\rm ph})=-v(\mathbf{q})q^2 {\rm Im}\,\pi_{\alpha\alpha}(\mathbf{q},\omega;T_{\rm el},T_{\rm ph})$, define the energy and corresponding damping (e.g., Landau damping and electron-phonon channel) of electronic excitations as a function of time delay (via time dependence of $T_{\rm el}$ and $T_{\rm ph}$). $v(\mathbf{q})$ is the Coulomb interaction and $\pi_{\alpha\alpha}$ is the current-current response function (photon self-energy) for the polarization direction $\alpha$. Useful optical quantities such as optical absorption $\sigma(\omega;T_{\rm el},T_{\rm ph})=i\pi_{\alpha\alpha}(\mathbf{q}=0,\omega)/\omega$ or photoconductivity $\Delta\sigma(\omega;T_{\rm el},T_{\rm ph})=\sigma(\omega;T_{\rm el},T_{\rm ph})-\sigma(\omega;T_{\rm el}=T_{\rm ph}=T_0)$ can also be simulated in the same way. The combination of Eq.\,\eqref{eq:epsilon_t} and the TTM equations was recently utilized in order to track ultrafast dynamics of Dirac plasmon in graphene, including energy loss of plasmon in non-equilibrium regime \cite{novko_first-principles_2021}. It was concluded that although steady state plasmon is mostly damped due to scatterings with acoustic phonons, the non-equilibrium Dirac plasmon dissipates most of its energy on intrinsic strongly coupled optical phonons. The same approach was also applied to study laser-induced band renormalizations close to van Hove singularity point of graphene band structure \cite{pagliara11,roberts14}, where it was concluded that electron-phonon coupling plays a significant role in photo-induced band modifications in graphene \cite{novko2019}.

Note that the approach of incorporating time dependence into Eqs.\,\eqref{eq:sigma} and \eqref{eq:epsilon_t} via the TTM results is actually general and can be applied to any spectral function or self-energy defined in terms of electron and phonon distribution functions. For instance, this approach was exploited to monitor hot-phonon dynamics in conventional superconductor MgB$_2$ via phonon spectral function $B(\omega;T_{\rm el},T_{\rm ph})$ and many-body phonon self-energy $\Pi(\omega;T_{\rm el},T_{\rm ph})$ of strongly-coupled $E_{2g}$ mode \cite{novko_ultrafast_2020}. All in all, this way of modelling ultrafast carrier and lattice thermalization can provide many microscopic insights of spectroscopic quantities in time domain, and was proven to have a well-balanced ratio between accuracy and numerical cost.

\begin{figure}[t]
\centering
\resizebox*{14cm}{!}{\includegraphics{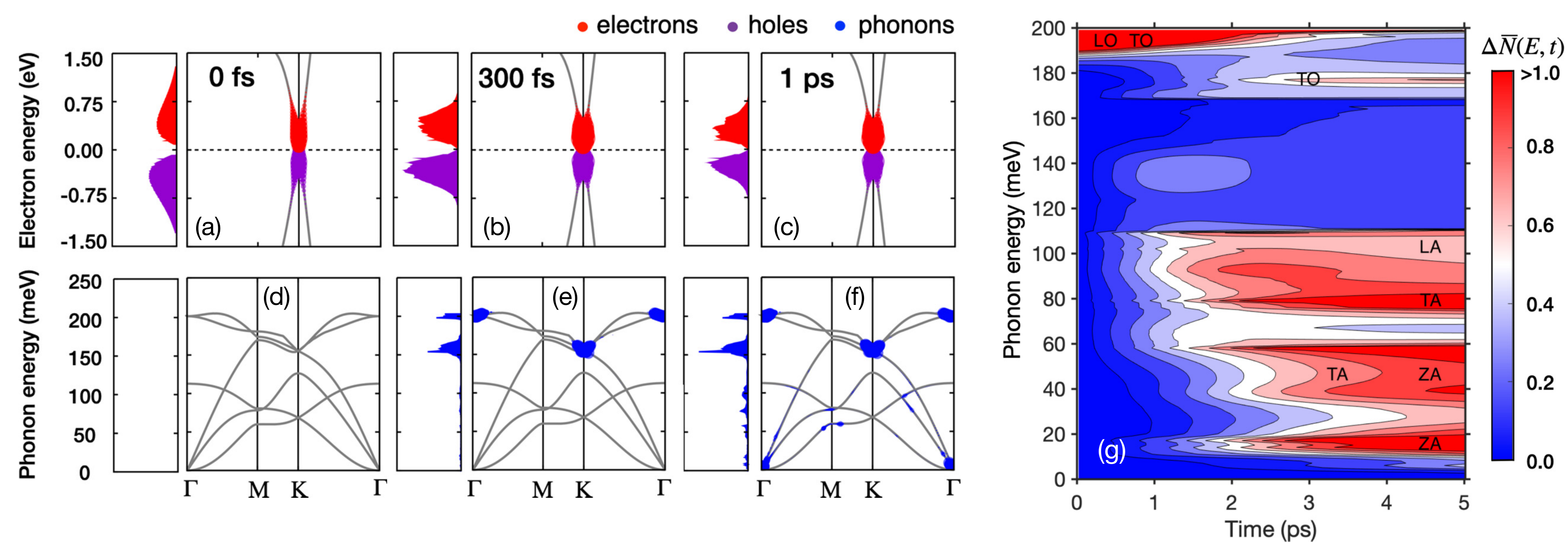}}
\caption{Population of excited electrons and holes for an initial electronic
excitation (a) superimposed to the band structure of graphene, and at
subsequent time steps throughout the non-equilibrium dynamics as obtained from
the solution of the time-dependent Boltzmann equation (b-c). The left panels
denote the density of states of electrons and holes.  (d-f) Time-dependent
phonon distribution functions $n_{\bq\nu}(t)$ superimposed to the phonon
dispersion for several time snapshot, with the point size being proportional to
$n_{\bq\nu}(t)$. 
(g) Energy resolved phonon population  $n_{\bq\nu}(t)$ as a function of time. 
Reproduced from Ref.~\cite{tong_toward_2021}.
} 
\label{fig:ber}
\end{figure}

\subsection{ Non-equilibrium lattice dynamics from the time-dependent Boltzmann equation}\label{sec:Cc}

To illustrate the application of the TDBE formalism to the ultrafast
electron-phonon dynamics of 2D materials and its suitability for the
description of these phenomena, we report in Fig.~\ref{fig:ber} ultrafast
dynamics simulations for graphene obtained from the numerical
solutions of Eqs.~\eqref{eq:bte_f} and \eqref{eq:bte_n} and reproduced from
Ref.~\cite{tong_toward_2021}. In panels (a)-(c), the electron ($f_{n\bk}$) and
hole (1-$f_{n\bk}$) distribution functions are superimposed to the band
dispersion of graphene for energies above and below the Fermi level (dashed)
for several time delays. The initial ($t=0$) distribution of electrons and
holes loses its excess energy via emitting phonon in the vicinity of the
$\Gamma$ and K high symmetry points and relaxes back to Fermi level within few 
picoseconds. The non-equilibrium phonon population is illustrated in
Figs.~\ref{fig:ber}~(d)-(f), where the point size is proportional to the phonon
occupation  $n_{\bq\nu}(t)$.  
The phonon population  at 300~fs (Fig.~\ref{fig:ber}~(e)) is characterized by 
an enhancement of the number LO and TO phonons at $\Gamma$
and K, indicating that these modes constitute the primary decay
path for excited electrons for the initial phases of the relaxation.  
These findings are compatible with the Eliashberg function shown in Fig.~\ref{fig:C}~(b). 
On longer time scales, the phonon-phonon scattering leads to the thermalization of
lattice and a redistribution to the excess energy of these modes to low-energy modes
Fig.~\ref{fig:ber}~(g). 

\begin{figure}[t]
\centering
\resizebox*{14cm}{!}{\includegraphics{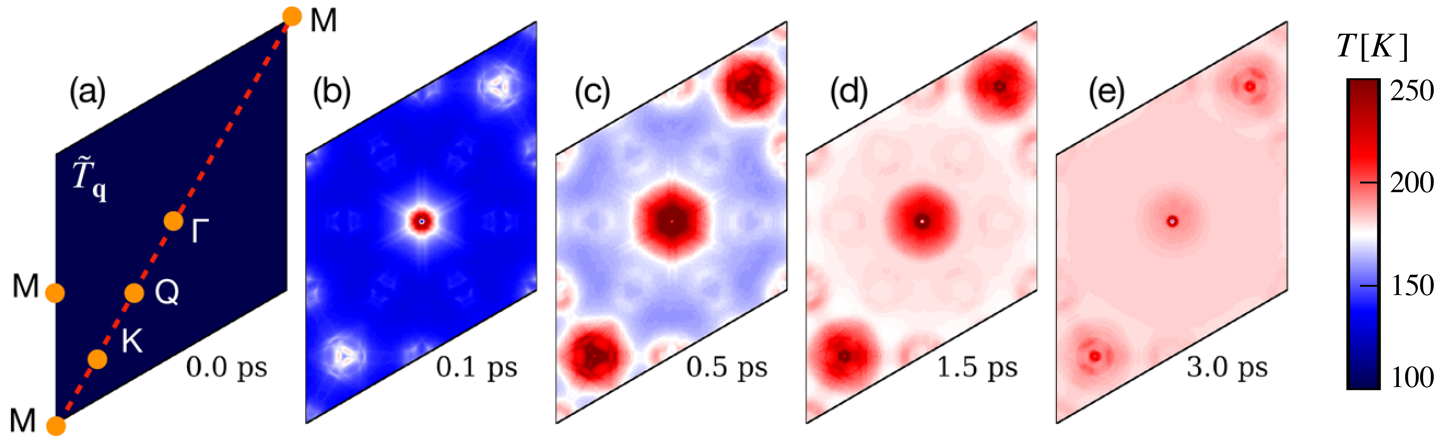}}
\caption{(a) Hexagonal Brillouin zone and high-symmetry points (dots) of 
monolayer MoS$_2$. The color coding reflects the effective vibrational temperature  
of the lattice for an initial ($t=0$) state of thermal equilibrium at $T=100$~K (see color bar). 
(b-e) Enhancement of the phonon temperature in the vicinity 
of the $\Gamma$ and K high-symmetry points due to momentum-selective  
phonon emission throughout the relaxation of a photo-excited electronic distribution. 
Reproduced from Ref.~\cite{caruso2021}.   }
\label{fig:jpcl}
\end{figure}

More generally, phonon-emission processes are constrained by energy and 
momentum conservation laws, which reduce the phase-space available for phonon-assisted 
electronic transitions. For example, graphene carriers in the vicinity of the 
Fermi level at K can scatter to states within the same Dirac cone at K, leading to the emission of 
phonons with ${\bf q}\simeq\Gamma$, alternatively they can scatter to the second Dirac cone at $-$K, emitting phonons 
with crystal momentum around K or -K. Transitions to other region of the BZ are forbidden as 
they violate  conservation laws. 
This stringent momentum selectivity confines the emitted phonons to narrow regions in 
reciprocal space, leading to the emergence of {\it hot spots} in the BZ, i.e., regions 
characterized by an enhanced vibrational temperature. 
Figure~\ref{fig:jpcl}~(a) illustrates the BZ and high-symmetry points of monolayer MoS$_2$, 
reproduced from Ref.~\cite{caruso2021}. 
The superimposed color coding denotes the effective vibrational temperature obtained by inverting the Bose-Einstein distribution 
as $ T_{\bq\nu}(t) = \hbar \omega_{\bq\nu} \{k_{\rm B} \ln [ 1 + n_{\bq\nu}(t) ]\}^{-1},$
and averaging over all mode indices $\nu$ via 
$\tilde  T_{\bq} =  N_{\rm ph}^{-1} \sum_\nu T_{\bq\nu}$ -- with
$N_{\rm ph}=9$ being the number of phonons of monolayer MoS$_2$. 
Here, the phonon distribution function $ n_{\bq\nu}(t)$ is obtained from the solution of the coupled 
TDBE for electrons and lattice. 
The preferential emission of phonons around $\Gamma$, K, and -$K$ is reflected by the 
enhanced vibrational temperature at these high-symmetry points for the initial phases of 
hot-carrier relaxation. Phonon hot-spots in the BZ can persist for several picoseconds, 
until the onset of phonon-phonon scattering processes restores a regime of thermal equilibrium. 

\section{Summary and Outlook}\label{sec:conc}

In this manuscript we introduced parameter-free theoretical approaches for the
study of the coupled ultrafast dynamics of electrons and phonons in solids, and
we reviewed their application to the prototypical two-dimensional materials, such as graphene and MoS$_2$.  

The two-temperature and the non-thermal lattice models
recast the thermalization dynamics of electrons and phonons into a set of
coupled first-order differential equations, which is formally equivalent to the
one governing the temperature evolution of coupled thermal baths.  While on the
one hand these approximations are sufficient to capture the characteristic
timescales and dynamics of the electron thermalization with the lattice
revealed by pump-probe experiments, on the other hand, the non-equilibrium electron,
vibrational and structural dynamics are poorly described, restricting the domain
of applicability of these approaches. In particular, the emergence of
non-equilibrium states of the lattice often entails phonon populations
characterized by strong anisotropy in the Brillouin zone that cannot be
captured by a Bose-Einstein function. 

The time-dependent Boltzmann equation overcomes these limitations.  As compared
to models, the advantage of this approach is two-fold: (i) it provides a
description of ultrafast electron and phonon dynamics with full momentum
resolution (ii) it accounts for phase-space constraints in phonon-assisted
electronic transitions. These points are important prerequisites to capture
the emergence of phonon hot-spots in the Brillouin zone, which have recently
been revealed by ultrafast diffuse scattering experiments
\cite{waldecker_eph_2016,Stern2018,rene_de_cotret_time-_2019,otto_mechanisms_2021,seiler_NL_2021}.
Additionally, the collision integrals can be made fully parameter-free by
determining electron-phonon and phonon-phonon coupling matrix elements via
available first-principles electronic-structure codes
\cite{Ponce2016,Perturbo,ShengBTE_2014}. 

Fostered by the relentless advances in the experimental characterization of
ultrafast phenomena, first-principles approaches for coupled
electron-phonon dynamics constitute a rapidly evolving research field. Many
non-equilibrium phenomena, however,  still represent {\it de facto} a challenge
for {ab-initio} techniques. 
These include, for instance, coherence and dephasing of electronic and
vibrational excitations \cite{marini2021coherence}, 
correlation effects on the electron dynamics \cite{Schl_nzen_2019},
structural phase transitions \cite{Juraschek2017,marini21,guan22}, topological Floquet states \cite{PhysRevResearch.2.043408},
and the transient formation of quasiparticle states \cite{Cotrete2113967119}. 
The development of accurate and efficient first-principles techniques capable of tackling
these phenomena remains a key research priority in the field of ultrafast science.
In this respect, approaches based on the density-matrix formalism \cite{winzer_carrier_2010} 
and non-equilibrium Green's functions \cite{Schl_nzen_2019} are promising routes to overcome the inherent limitations of the Boltzmann
equation: recent studies explored new directions to 
reduce the notoriously high computational cost of NEGF methods 
\cite{Bonitz2020} paving the
way towards a rigorous quantum mechanical description of coherent and
correlated dynamical phenomena in condensed matter \cite{Perfetto2022}.

\section*{Acknowledgement(s)}
This project has been funded by the
Deutsche Forschungsgemeinschaft (DFG) -- Projektnummer 443988403.
D.N. acknowledges financial support from the Croatian Science Foundation (Grant no. UIP-2019-04-6869) and from the European Regional Development Fund for the ``Center of Excellence for Advanced Materials and Sensing Devices'' (Grant No. KK.01.1.1.01.0001).

\noindent\textbf{Appendix A. Analytical solution of the TTM}\medskip

Below we report the analytical solution of the
TTM for the limiting case of temperature-independent heat capacities. The TTM equations in absence of external driving field are reported below: 
\begin{align}
 \frac{\D T_{\rm ph}}{\D t} &= \frac{g}{C_{\rm ph}}( { T_{\rm el}- T_{\rm ph} }) \quad, \\ 
 \frac{ \D T_{\rm el}}{\D t }
&= \frac{g} {C_{\rm el}}  ({ T_{\rm ph}- T_{\rm el}})  \quad.
\end{align}
The TTM thus forms a set of two coupled first-order differential equation, which can be solved analytically via standard techniques. Its solution is: 
\begin{align}
  T_{\rm el} (t) &= a_1 e^{-\gamma t} + a_2  \quad,\\ 
  T_{\rm ph} (t) &= - \frac{C_{\rm el}a_1}{C_{\rm ph}} e^{-\gamma t} + a_2 \quad,
\end{align}
where we introduced the constants 
$a_1 = (T_{\rm el}^0 - T_{\rm ph}^0) {C_{\rm ph}}/( {C_{\rm ph}}+{C_{\rm el}})  $,  
$a_2 = T_{\rm el}^0 -c_1 $,
and $\gamma = g/(1/C_{\rm el} + 1/C_{\rm ph})$. 
The solution is easily verified by substitution. 

By introducing a source term coupled to the electronic temperature, the TTM gets modified as follows:
\begin{align}
 \frac{\D T_{\rm el}}{\D t} &= \frac{g}{C_{\rm el}}( { T_{\rm ph}- T_{el} })  + S(t) \quad,
\end{align}
where $S(t)$ describes the interaction with a light pulse, and the second equation remains unchanged. 
Analytic solution of the TTM is still possible for a simple time dependence of $S(t)$. 
For example, by considering $S(t)=\alpha e ^{-\frac{t}{\tau}}\theta(t)$, the solution can be written as: 
\begin{align}
  T_{\rm el} (t) &= -\tau b_1 e ^{-\frac{t}{\tau} } - b_2 \gamma^{-1} e^{-\gamma t} + b_3  \quad,\\ 
  T_{\rm ph} (t) &=   T_{\rm el} (t) + \frac{C_{\rm el}}{g} \left[ (b_1-\alpha)  e ^{-\frac{t}{\tau} } + b_2    e^{-\gamma t} \right] \quad,
\end{align}
with the constants: 
\begin{align}
  b_1 &= \alpha \left( \frac{g\tau}{ C_{\rm ph}} - 1\right)( \gamma\tau-{1} )^{-1} \quad,\\ 
  b_2 &= \frac{g}{C_{\rm el} } (  T_{\rm ph}^0 -   T_{\rm el}^0) + \alpha -b_1  \quad, \\
  b_3 &=  T_{\rm el}^0 +  \frac{b_2}{\gamma} + b_1 \tau \quad.
\end{align}


\end{document}